\documentclass[11pt,a4paper]{article}
\usepackage[utf8]{inputenc}
\usepackage[T1]{fontenc}
\usepackage{lmodern}
\usepackage[margin=2.5cm]{geometry}
\usepackage{amsmath,amssymb}
\usepackage{graphicx}
\usepackage{setspace}
\usepackage[font=small,labelfont=bf]{caption}
\usepackage[hidelinks]{hyperref}
\usepackage{xurl}
\usepackage{xcolor}

\graphicspath{{figs/}}
\hypersetup{
  pdftitle={Environmental seismology: What can we learn on earth surface processes with ambient noise?},
  pdfauthor={E. Larose, S. Carrière, C. Voisin, P. Bottelin, L. Baillet, P. Guéguen, F. Walter, D. Jongmans, B. Guillier, S. Garambois, F. Gimbert, C. Massey},
  pdfsubject={Preprint},
  pdfkeywords={Environmental seismology, Engineering seismology, Ambient noise, Hydrology, Thermo-mechanics}
}

\newcommand{\dvv}{dV/V}

\begin{document}

\begin{center}
\fbox{\parbox{0.92\textwidth}{\small\setstretch{1.1}
\textbf{Preprint.} This is the authors' submitted version of the article published in
\emph{Journal of Applied Geophysics}, 116 (2015), 62--74.
The final version is available at \url{https://doi.org/10.1016/j.jappgeo.2015.02.001}.
This version may differ from the published version.}}
\end{center}

\vspace{1em}
\begin{center}
{\Large\bfseries Environmental seismology: What can we learn on earth surface processes with ambient noise?\par}
\vspace{1.2em}
{\large Eric Larose$^{a,b,c,*}$, Simon Carrière$^{a,b,c}$, Christophe Voisin$^{a,b,c}$, Pierre Bottelin$^{a,b,c}$, Laurent Baillet$^{a,b,c}$, Philippe Guéguen$^{a,b,c}$, Fabian Walter$^{a,b,c}$, Denis Jongmans$^{a,b,c}$, Bertrand Guillier$^{a,b,c}$, Stéphane Garambois$^{a,b,c}$, Florent Gimbert$^{d,e}$, Chris Massey$^{f}$\par}
\vspace{0.8em}
{\small\setstretch{1.1}
$^{a}$ISTerre, CNRS, BP 53, 38041 Grenoble Cedex 9, France\\
$^{b}$ISTerre, Univ. de Grenoble Alpes, BP 53, 38041 Grenoble Cedex 9, France\\
$^{c}$ISTerre, IFSTTAR, BP 53, 38041 Grenoble Cedex 9, France\\
$^{d}$Seismological Laboratory, California Institute of Technology, Pasadena, CA, USA\\
$^{e}$Division of Geological and Planetary Sciences, California Institute of Technology, Pasadena, CA, USA\\
$^{f}$GNS Science, Lower Hutt 5040, New Zealand\\
$^{*}$Corresponding author: eric.larose@ujf-grenoble.fr\par}
\end{center}

\section*{Abstract}
Environmental seismology consists in studying the mechanical vibrations that originate from, or that have been affected by external causes, that is to say causes outside the solid Earth. This includes for instance the coupling between the solid Earth and the cryosphere, or the hydrosphere, the anthroposphere and the specific sources of vibration developing there. Environmental seismology also addresses the modifications of the wave propagation due to environmental forcing such as temperature and hydrology. Recent developments in data processing, together with increasing computational power and sensor concentration have led to original observations that allow for the development of this new field of seismology. In this article, we will particularly review how we can track and interpret tiny changes in the subsurface of the Earth related to external changes from modifications of the seismic wave propagation, with application to geomechanics, hydrology, and natural hazard. We will particularly demonstrate that, using ambient noise, we can track 1) thermal variations in the subsoil, in buildings or in rock columns; 2) the temporal and spatial evolution of a water table; 3) the evolution of the rigidity of the soil constituting a landslide, and especially the drop of rigidity preceding a failure event.

\medskip
\noindent\textbf{Keywords:} Environmental seismology; Engineering seismology; Ambient noise; Hydrology; Thermo-mechanics

\section{What is environmental seismology?}

Seismology has long been considered to be the primary source of information on the structure of the Earth, and deep Earth studies have raised a broad scientific interest. From the arrival times, amplitudes, polarizations of refracted and reflected waves, one can deduce the structure of the Earth, which relates to its geological composition (Astiz et al., 1996; Shearer, 2009). Based on these observations, seismologists have produced 1D, 2D and 3D cartographies of the Earth at various scales. At the global scale, very deep and crustal studies offered key indications for the geodynamics of the Earth's interior. At kilometric scale, seismology yields to industrial application in reservoir exploration and characterization, and at hectometric to metric scales seismology offer application to geotechnical and civil engineering, including landslides and soil characterization.

Recent progresses in hardware technology (sensor sensitivity, timing, concentration), computational power (massive data mining), and methodology (use of ambient noise correlation, coda waves, data mining, advanced signal processing) now allow observing of new kinds of seismic waveforms obtained from continuous records. From these new datasets, one can now track very tiny changes in the subsurface, or unveil very small and unconventional seismic sources out of the ambient noise, yielding to the development of a new field in seismology: environmental seismology.

Environmental seismology consists in studying natural seismic vibrations that are either triggered by processes occurring outside of the solid Earth (the cryosphere, hydrosphere, atmosphere and beyond\ldots), or whose propagation in the solid Earth is perturbed by modifications of environmental external parameters (temperature, hydrology\ldots) or human activity. Generally speaking, environmental seismology splits into two fields. The first field aims at studying the modification of wave propagation in the solid earth due to processes in relation with the external environment, including hydro-meteorological phenomena, thermal evolution, and erosion processes. The second field of environmental seismology concerns the study of natural seismic sources that are triggered by these external phenomena, including those developing in the atmosphere (wind, storms\ldots), the cryosphere (ice quakes), and the hydrosphere (river noise, ocean hum\ldots). The present paper proposes to touch on these very broad research fields where seismology has recently offered new perspectives, and which were not accessible to traditional seismological techniques a few years ago.

\begin{figure}[htbp]
\centering
\includegraphics[width=\textwidth]{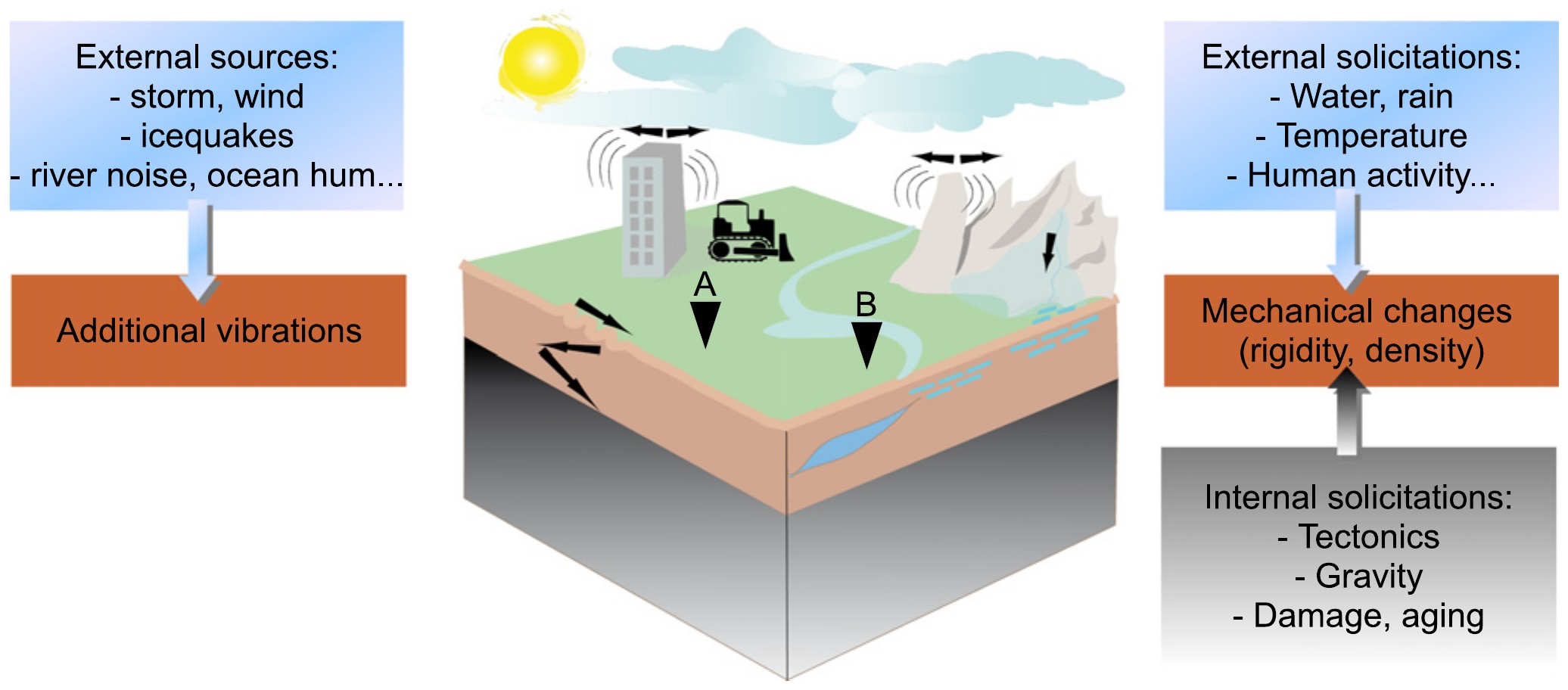}
\caption{The Earth subsurface is subject to various mechanical solicitations, which likely change the seismic waveforms. On the left: sources of seismic vibrations originating from external processes. On the right: example of mechanical changes either induced by environmental changes, or induced by telluric phenomena.}
\label{fig:sketch}
\end{figure}

In order to monitor a medium with seismology, one needs a reproducible seismic signal propagating in the medium. This used to be exclusively achievable with active sources (explosives, hammer blows, etc.) or moderate repetitive earthquakes, but over recent years people have figured out how to use passive recordings, only. These new techniques use ambient noise records at seismometer pairs to reconstruct the seismic response at one receiver as if an active source had been placed at the other. With this new approach, one can track for example tiny changes in the Earth, especially those due to external perturbations (see Fig.~\ref{fig:sketch}). These perturbations can be due to human activity (oil, gas and geothermal exploitation, mining, urbanization), environmental coupling (hydrology, temperature) or internal forcing (gravity, tectonics). The main goal of the present paper is two-fold: to identify and quantify external forcing applied to the subsurface, with two objectives: first, learn about environmental processes; and second, discriminate the effects of environmental forcing on seismic waveforms from the effects of internal changes (state of stress, change in rheology, damage and fracturing\ldots). An important issue is also to discriminate reversible changes from irreversible ones.

Environmental seismology also consists in digging new signals out of the ambient noise. Thanks to their frequency signature, wave polarization, amplitude and duration, these signals sometimes reveal new kinds of seismic sources, such as those observed in glaciers or on the ice shelf, from river noise, and from the ocean. The discovery of these sources shed a new light on the physical processes at work in these natural objects.

In the first section of this article, we recall the basics of processing ambient seismic noise to obtain reproducible correlograms, and the subsequent processing to track tiny changes in the material. These changes are, in Section~\ref{sec:thermal}, analyzed in terms of thermal variations, and in Section~\ref{sec:hydro} they are related to hydrological loading. In Section~\ref{sec:landslide}, we present an application to monitoring landslide activity, and exhibit a candidate for landslide precursor. In Section~\ref{sec:sources}, we briefly review recent observations of seismic sources in the ice and in rivers. The formers give new insight into glacier deformation or ice shelf evolution. The latter yields indications about river flow and solid transport.

\section{Processing ambient seismic noise}

\subsection{Ambient noise spectrum}

One of the very common features of ambient noise that any seismologist can exploit is the frequency content of the record. This frequency content depends on the spectrum illuminated by the sources, which in practice cover all the seismic frequencies from a few milli-Hertz to several hundreds of Hertz, but also depends on the attenuation of the waves along their trajectories, and on local effects such as structural and geometrical structuration of the sub-surface. The power spectrum density is simply derived from the continuous records by averaging the intensity of the fast Fourier transform (FFT) of the record: $PSD(f) = |FFT(\varphi_A(t))|^2$. Depending on the geometry of the geological structures, the power spectrum density can be larger at some specific frequencies. This is the case, for instance, in resonant structures such as sedimentary basins, mountain edges, rock columns, or even buildings. For the last two examples, the simple model of a bending beam of density $\rho$, vertical elevation $L$ and square cross-section $S$ yields the resonant frequency $f \propto 1/L^2 \sqrt{ES/\rho}$, where $E$ is the Young's modulus (Humar, 1990). This means that the resonant frequency decreases with the mass and the length of the structure, and increases with its rigidity. Following the resonant frequency of a given structure thus allows us to give indications on the evolution of either the geometry of the structure (size of a rock column), its mass, or the quality of the material (evolution of its rigidity).

\subsection{Ambient noise correlation}

\begin{figure}[htbp]
\centering
\includegraphics[width=0.6\textwidth]{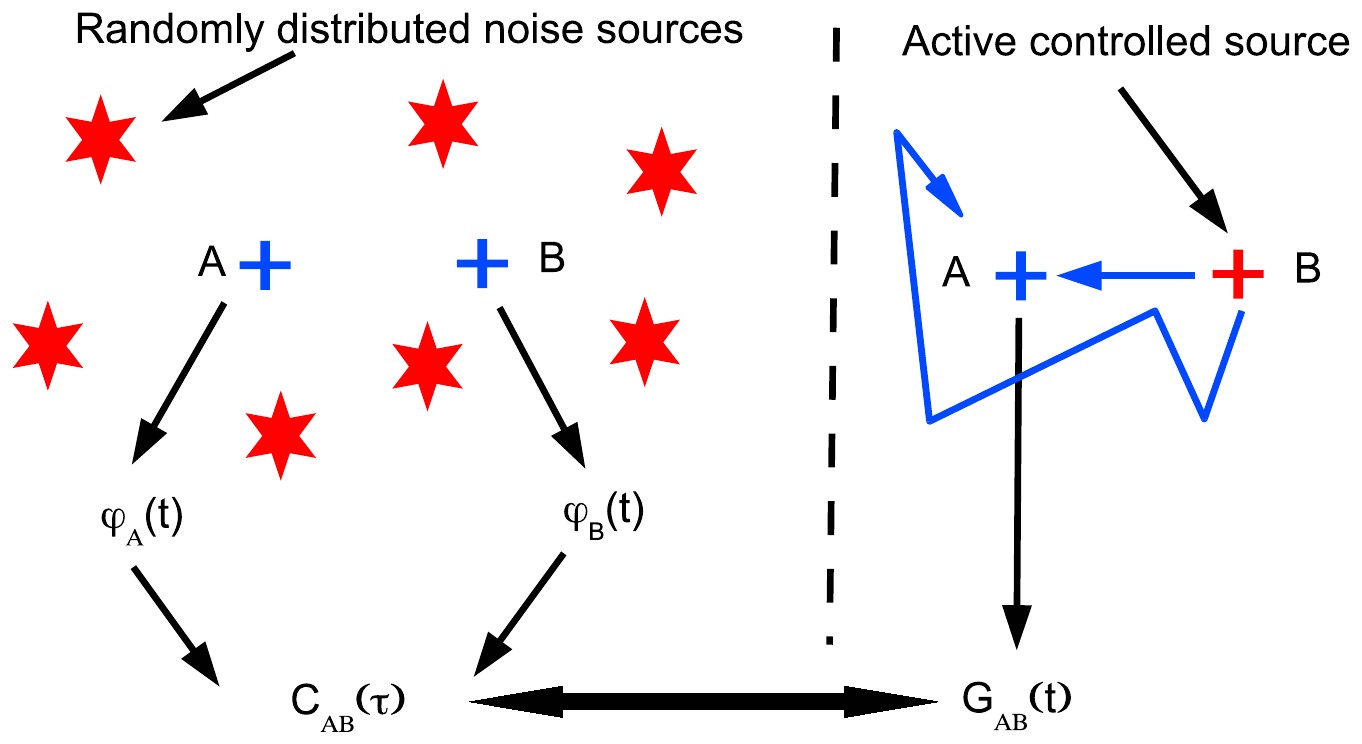}
\caption{Conceptual equivalence between passive experiment based on ambient noise correlation (left) and active experiment (right). Red stars: noise sources; blue crosses: receivers; red cross: controlled source. The correlation $C_{AB}(\tau)$ between the two simultaneous records $\varphi_A(t)$ and $\varphi_B(t)$ yields the impulse response $G_{AB}(\tau)$ provided that enough noise sources surround the receivers. Blue arrows: direct and multiply scattered (coda) waves.}
\label{fig:concept}
\end{figure}

The retrieval of the seismological response of the subsurface from noise is a field that has experienced explosive growth over the past decade. The main idea is that, by using the time-correlation $C(\tau)$ performed on signals $\varphi(t)$ recorded at the same time at two different places A and B, one has access to the impulse response $G(t)$ between A and B as if one of the receiver was a source. We summarize this concept in Fig.~\ref{fig:concept}, where we illustrate that the cross-correlation between A and B yields the Green's function provided that enough noise sources are distributed around the sensors. The exact relation reads:
\begin{equation}
\partial_\tau C_{AB}(\tau) = \partial_\tau \int \varphi_A(t)\,\varphi_B(t+\tau)\,dt \;\propto\; G^+(A,B,\tau) - G^-(A,B,-\tau),
\label{eq:corr}
\end{equation}
where $G^+$ and $G^-$ stand for the causal and anti-causal Green's function, respectively. This approach offers the opportunity to mimic sources almost everywhere we are able to place a seismic sensor, and to actually turn noise into signal.

To be more precise, the conditions to meet for the correlations to converge to the Green's function are (Curtis et al., 2006; Sato, 2009):
\begin{itemize}
\item sources are uncorrelated in time
\item to reconstruct surface waves, sources are located all around the receivers,
\item to reconstruct bulk waves,
\item the wave field is equipartitioned, meaning a proper ratio of compressional and shear waves.
\end{itemize}
In practice, the last three conditions are hardly met in nature. Fortunately, scattering of waves, which implies mode conversion and scattering in all directions, compensate for the lack of equipartition in modes and propagation directions (Weaver and Lobkis, 2001; Paul et al., 2005; Larose et al., 2005).

The technique of extracting the impulse response from noise measurements is known by different names that include Green's function retrieval, seismic interferometry, passive imaging through ambient noise correlation, and others. The ability to use ambient noise as a source was initially proposed in helioseismology (Duvall et al., 1993), a concept that was generalized by the fertile paper of Weaver and Lobkis (2001) to other fields like seismology (Campillo and Paul, 2003; Shapiro and Campillo, 2004), ocean acoustics (Roux et al., 2004; Sabra et al., 2005), ultrasound (Derode et al., 2003; Larose et al., 2006a), and structural engineering (Snieder and Safak, 2006), amongst many others. We attract the reader's attention on review articles such as Larose et al. (2006b), Curtis et al. (2006), Campillo et al. (2011), and Snieder and Larose (2013) for more detailed comments and description on the concept, and a more detailed literature.

\subsection{Monitoring with ambient noise correlation}
\label{sec:monitoring}

As ambient noise is continuously available on Earth, we have the possibility to reconstruct the impulse response between sensors at different calendar dates. Sens-Schönfelder and Wegler (2006) proposed to compare correlograms from one day to another (or compare a daily correlograms to a reference averaged over a much larger period) to monitor tiny velocity changes in a volcano. This idea derived from older studies that either used repetitive earthquakes (called doublets) (Poupinet et al., 1984) or active reproducible sources (Snieder et al., 2002), but now such sources are not required. For time-lapse monitoring of Earth properties, getting rid of sources issues is certainly a major opportunity!

Provided that the sensors are fixed, and that the ambient noise sources are relatively stable from a statistical point of view, changes in the waveforms of the correlograms are solely due to mechanical changes in the subsurface. Actually, laboratory experiments (Hadziioannou et al., 2009) also demonstrated that even if the noise structure differs from one day to another, and that the Green's function reconstruction is imperfect in the correlation, passive time-lapse monitoring (also called Passive Imaging Interferometry, or Passive Coda Wave Interferometry) still yields the proper mechanical change within the medium, the necessary condition being that at least part of the noise sources should remain stable along calendar dates.

\begin{figure}[htbp]
\centering
\includegraphics[width=0.6\textwidth]{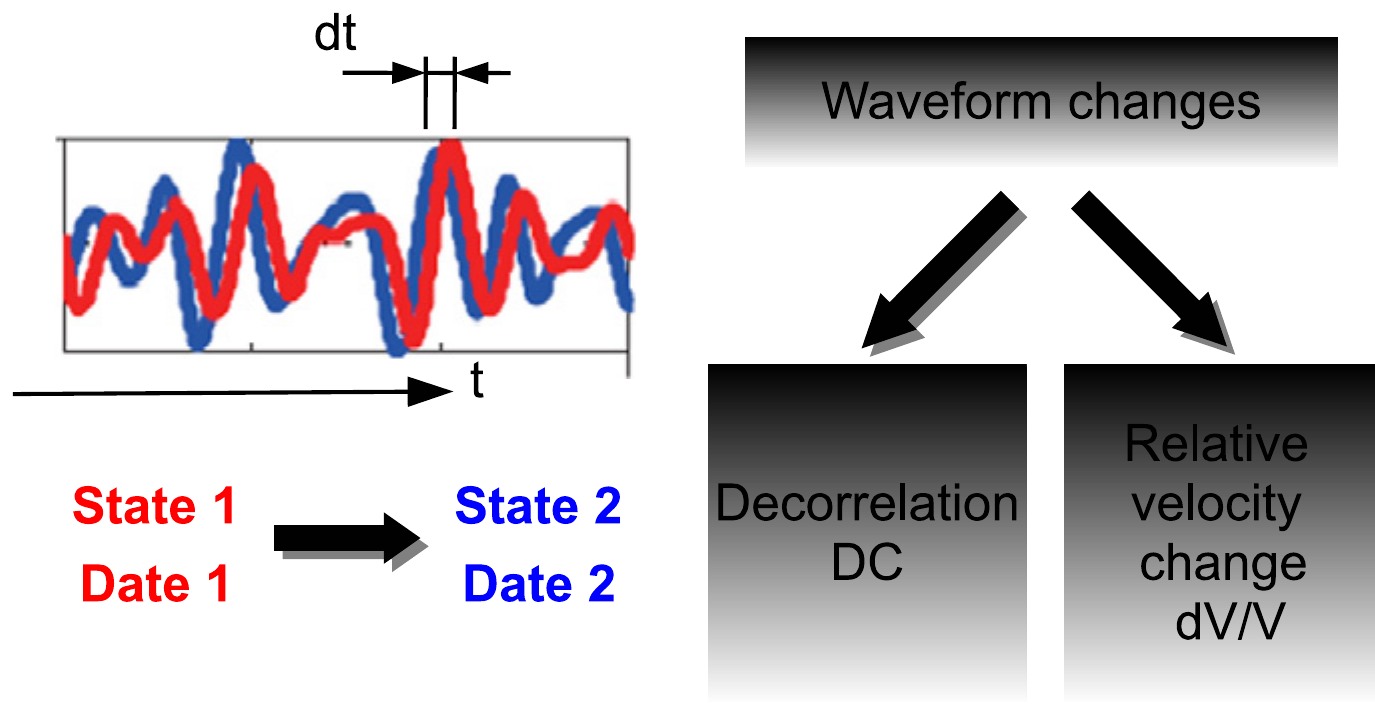}
\caption{Example of waveform change on synthetic seismograms. Red and blue waveforms are acquired at the same place but at different dates, and differ from a slight delay $dt$ at time $t$ in the record, which results from a relative velocity change in the material $\dvv = -dt/t$. After correcting from this time delay, the waveforms still do not exactly fit, which is quantified by the decorrelation coefficient DC. This decorrelation results from geometrical or structural changes in the subsurface.}
\label{fig:waveform}
\end{figure}

Let's now focus on the nature of the mechanical changes that we can monitor, and how they materialize in the seismic waveforms. The simplest observation is certainly a change in arrival time $dt$ for a given wave-packet (see Fig.~\ref{fig:waveform}) arriving at time $t$. This wave-packet can be either a direct wave, or late arrivals, but in practice we prefer to analyze the latter, also called coda waves, as they are more stable against changes in ambient noise source locations and much more sensitive to mechanical variations of the subsurface. The intuitive reason for the high sensitivity of diffuse waves is simply that, as they spend much more time traveling in the material, they accumulate more changes along their paths than direct waves (but are also much harder to interpret). This arrival time change results from a change in the seismic wave velocity, and if the material undergoes a homogeneous velocity change, then we have the simple relation: $\dvv = -dt/t$. The records $\varphi(t)$ thus undergo a simple stretching of the time axis:
\begin{equation}
\varphi(t) \rightarrow \varphi\left(t\,(1 - \dvv)\right).
\label{eq:stretch}
\end{equation}
We can then correct the waveforms from this relative velocity change, and check if the waveforms are perfectly matching. In general, even after this correction, waveforms still show a slight decorrelation DC, which can be interpreted as structural or geometrical changes within the material. In other words, the positions of the structures were slightly modified. This is for instance the case when the monitored medium undergoes fluid migrations, local deformation etc. Changes in amplitude or polarization are other quantities that can be derived from comparing the waveforms, but are more challenging to observe with significant reliability when using ambient noise correlation techniques.

\begin{figure}[htbp]
\centering
\includegraphics[width=\textwidth]{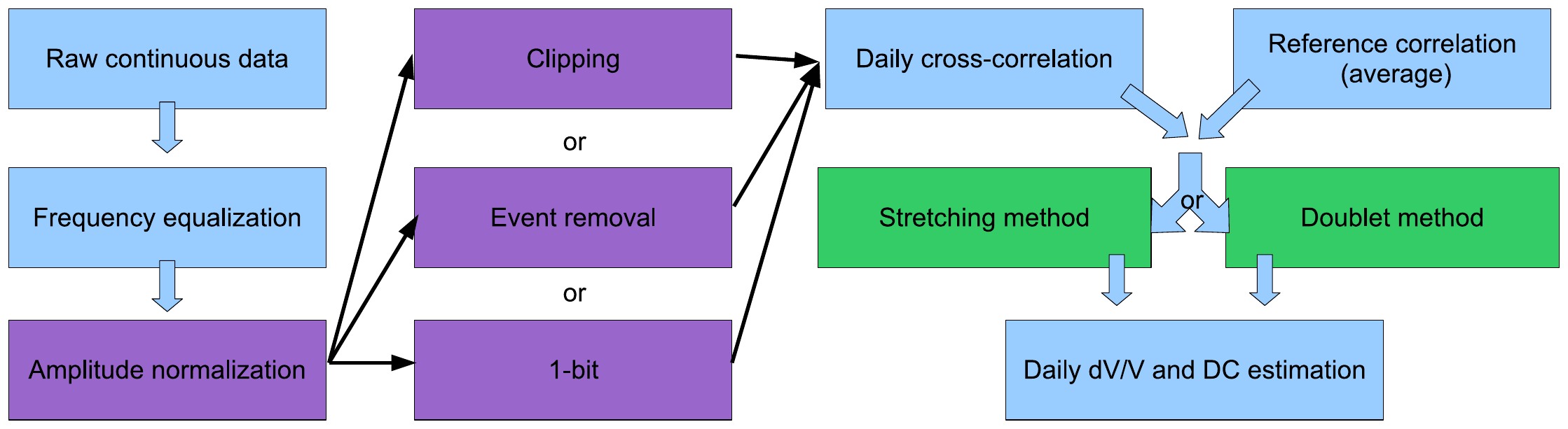}
\caption{General workflow to process ambient seismic noise correlation and extract daily relative velocity changes $\dvv$ and decorrelation values DC.}
\label{fig:workflow}
\end{figure}

We now briefly recall the general workflow to obtain daily relative velocity changes $\dvv$, and decorrelation values DC, from ambient noise collected at two passive sensors (see Fig.~\ref{fig:workflow}). The first step is to equalize the frequency content of the ambient noise (frequency normalization, or spectral whitening) in the frequency domain. Then, in order to attenuate the statistical weight of rare but large amplitude events, we perform an amplitude normalization in the time domain (see purple boxes in Fig.~\ref{fig:workflow}). There are different ways to perform such an amplitude normalization, which includes 1-bit processing (Campillo and Paul, 2003; Larose et al., 2004), event removal, or amplitude clipping (Bensen et al., 2007). In 1-bit processing, one only retains the sign (+1 or $-1$) of the waveform before cross-correlation, a fast and simple procedure that attracted much attention at the beginning, but that seems less efficient than the clipping procedure (limitations due to information degradation, and non-linear transformation of the data). Clipping consists in saturating the waveforms for amplitudes larger than a few times the rms value of the record, therefore it reduces the weight of large events while keeping all the pieces of information contained in the low-amplitude continuous noise. The readers could refer to those previous articles for technical details on the amplitude normalization. Then, from the set of daily correlations, we define a reference function, for instance by averaging the correlations over all the available dates. Each daily correlation $C^d_{AB}(t)$ is subsequently compared to the reference $C^{ref}_{AB}(t)$. Again, there exist in the literature at least two strategies to extract the relative velocity changes: the doublet method (Poupinet et al., 1984), and the stretching method (Lobkis and Weaver, 2003; Sens-Schönfelder and Wegler, 2006; Hadziioannou et al., 2009). Both techniques have advantages and drawbacks. Briefly speaking, the doublet technique has the advantage of giving an estimation of the change at a given time in the coda, but has limited stability, whereas the stretching technique is more stable, but the measurement of the change is averaged over a larger time-window. The main reason here to prefer the stretching approach is that it directly gives access to the decorrelation values DC, together with the daily relative velocity change $\dvv$. Note that we can also estimate the error around a given value of $\dvv$ from the value of DC, as proposed by Weaver et al. (2011).

The next section expands on the different geophysical sources of $\dvv$ and DC changes observed in the seismic waveforms.

\section{Thermal forcing}
\label{sec:thermal}

\subsection{Theoretical and experimental background}

Seismic or acoustic wave velocities in solids and fluids are well known to vary with temperature. For instance water has a thermal relative velocity change coefficient $\dvv/dT$ of $+3.3 \times 10^{-3}$/\textdegree C, and dry air $+2 \times 10^{-3}$/\textdegree C, which represents an increase of velocity with increasing temperature. In rocks and other aggregates like concrete, this coefficient has negative values of the order of $-2 \times 10^{-3}$/\textdegree C (Snieder et al., 2002; Larose et al., 2006c).

The subsurface undergoes at least two kinds of thermal forcing: daily and yearly variations. On the short term, the outside air temperature variation is of the order of $\pm$10\,\textdegree C over 24~h, and has a limited depth penetration such that, at one meter depth, the daily variation is of the order of 0.1\,\textdegree C or less. Yearly variations penetrate deeper such that they are of the order of $\pm$10\,\textdegree C at 1~m and reduce down to $\pm$1\,\textdegree C at 5~m depth (Tsai, 2011). These figures are only given in order of magnitude and can significantly vary from one site to another, but we can draw the general idea that thermal variations of the soil can induce only a feeble relative seismic wave velocity change, as low as $10^{-3}$ or less, in the first meters, and no velocity change at all at depth greater than a few tens of meters. Note that this approach does not take into account stress variations associated to differential dilation of the material (Tsai, 2011).

In most conventional seismic field experiments, we do not measure the effect of external thermal changes on seismic attributes (amplitude, arrival time, polarization\ldots). The first reason probably relates to the limited effect of external temperature changes on the soil, as mentioned in the previous paragraph: the depth penetration of those variations hardly goes beyond a few meters at most, thus only a very thin and shallow layer might be subject to thermal changes. Another reason might be found on the technological side: due to the limited frequency band at work, together with the electronic and seismic background noise, the precision of the arrival time measurement is often too limited to detect variations in relative velocities as low as $10^{-3}$ to $10^{-4}$.

A standard technique to measure the effect of temperature on wave velocities, or in general on the elastic moduli of the material, is to evaluate the change of arrival time of the wave packet launched at a fixed source and received at a fixed sensor. Coda Wave Interferometry (CWI), a technique that takes benefit of late arrivals named coda waves, is an alternative procedure that has demonstrated very high sensitivity to weak variations (Poupinet et al., 1984; Snieder et al., 2002). Active CWI experiments have been performed to monitor the temperature with elastic waves in the laboratory for rock physics (Snieder et al., 2002), civil engineering (Larose et al., 2006c), and acoustics (Planes and Larose, 2013). Several obstacles limit the applicability of CWI to seismology: first, it is very hard to operate a reproducible source; second, obtaining coda waves necessitates employing powerful sources. Sens-Schönfelder and Wegler (2006) proposed to use ambient seismic noise correlations to produce daily, weekly or monthly seismic waveforms. The stability of the correlograms allows performing CWI without any active and controlled source. This is the idea that we already developed in Section~\ref{sec:monitoring}.

\subsection{Application to the subsurface}

\begin{figure}[htbp]
\centering
\includegraphics[width=0.6\textwidth]{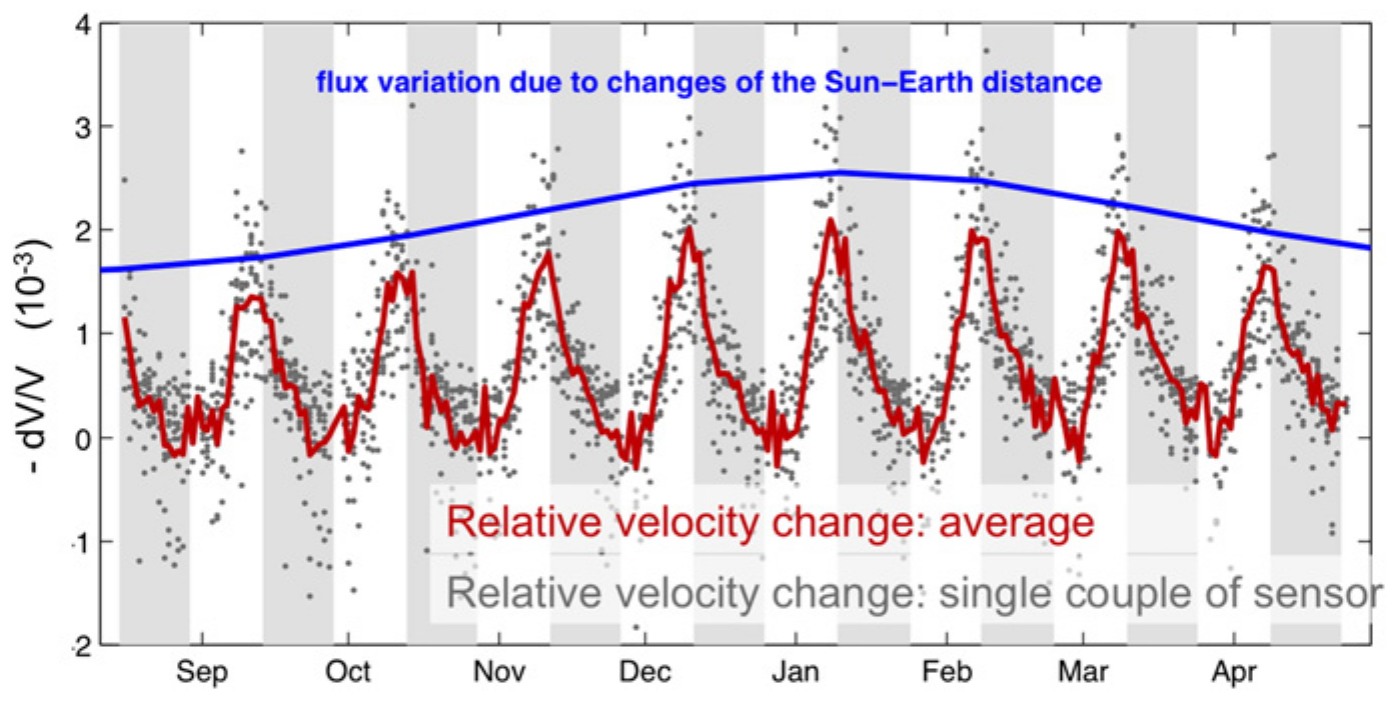}
\caption{Relative seismic velocity variations monitored on the moon at the Apollo 17 landing site, in red, and sun--moon distance in blue. Data have been collected from 1976 to 1977. This figure is derived from Sens-Schönfelder and Larose (2008).}
\label{fig:moon}
\end{figure}

Giving an example of thermal monitoring of the subsurface with ambient seismic vibration is not trivial, as the subsurface often undergoes several solicitations at the same time. In an ideal experiment, we would change only the temperature, with no weather activity, moisture evolution, or human activity. This dream is hardly met on Earth, but is actually possible thanks to the incredible seismic dataset collected during the Apollo era on the moon. A few years ago, Sens-Schönfelder and Larose (2008) reprocessed an old dataset using the modern methodology proposed earlier in this paper (see Fig.~\ref{fig:workflow}). The data were collected continuously from four geophones (array aperture of the order of 100~m), from 1976 to 1977, at the Apollo 17 landing site. The frequencies used in the study are in the standard range of surface wave seismic prospection (a few Hertz to a few tens of Hertz). In Fig.~\ref{fig:moon} we plot the relative velocity change observed every (terrestrial) day for each couple of receivers (gray dots) and averaged over the array (red solid line). $\dvv$ demonstrates fluctuations of one lunar day period (29.5 terrestrial days) that can only be explained by solar radiation heating: seismic velocity is found to decrease with increasing temperature at daytime (white background, note the negative value of the y-axis), and increases with decreasing temperature at nighttime (gray background). Interestingly, the absolute amplitude of these fluctuations coincides perfectly with the evolution of the sun--moon distance, which can be simply explained by the total radiation intercepted by the moon, and which depends on its distance from the sun along the year. These results demonstrate the possibility to monitor thermal changes in the soil from ambient seismic noise.

\subsection{Application to civil engineering}

Ambient vibration based methods provide an effective solution for short- and long-term monitoring of buildings. The basic idea is that changes of stiffness impact the modal frequencies of a system and modify its dynamic response (Doebling et al., 1996; Farrar and Worden, 2007). In actual buildings, the variation of the fundamental frequency can be due to changes in the boundary conditions between the soil and structure, changes in the design properties such as retrofitting, or changes in the elastic properties of the material. Permanent variations can appear due to structural damage caused by repetitive strong seismic motion (e.g., Clinton et al., 2006; Michel and Gueguen, 2010). But recently, thanks to the emergence of new data (high sensitive- and continuous recordings), long-term variations have been observed, and were found to be reversible and of limited amplitude. In most cases, these fluctuations are related to the temporal variations of the atmospheric conditions (such as temperature and humidity) that may influence the properties of the structure and the boundary conditions.

\begin{figure}[htbp]
\centering
\includegraphics[width=0.6\textwidth]{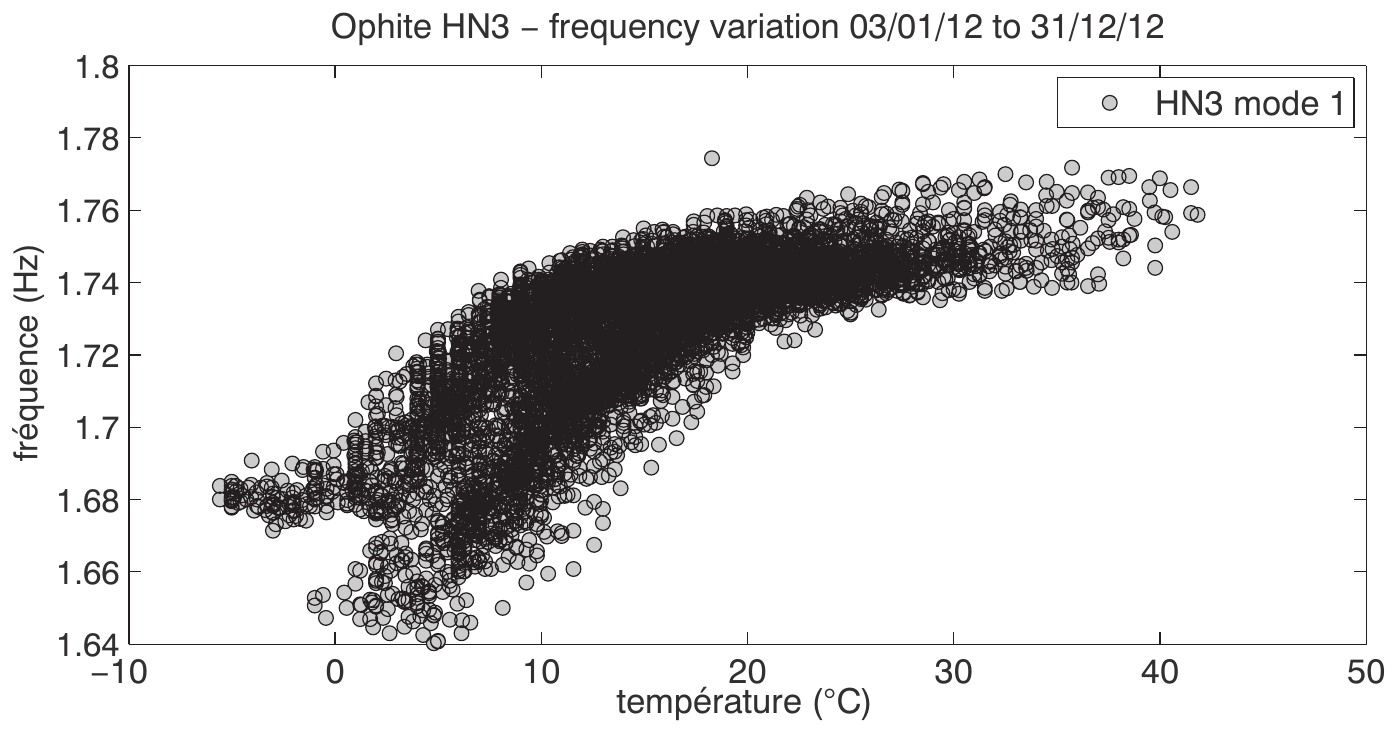}
\caption{Relation between the resonant frequency of the Ophite tower building and the outside air temperature in the transverse (HN3) direction. Only the first mode is displayed. Each dot corresponds to the frequency of the fundamental mode extracted using the Random Decrement Technique on window for 1~h applied to one year of data collected in 2012. Data were processed similarly to those collected in 2010 and published in ref. Mikael et al. (2013).}
\label{fig:ophite}
\end{figure}

Fig.~\ref{fig:ophite} shows the correlation of the fundamental frequency of the Ophite tower (France) with temperature. The Ophite Tower is an 18-story dwelling building built in 1972. The lateral resistance system in the two horizontal directions is provided by reinforced concrete shear walls. Its dimensions are 19-m wide, and 24-m long. This building is founded on a rocky site, composed of Ophite rock, and we therefore assume a shallow foundation system. Since 2010, the Ophite Tower has a permanent monitoring launched by the French Accelerometric Network (Péquegnat et al., 2008). Twenty-four accelerometric channels were installed recording continuously the building vibrations in the two horizontal (HN2 and HN3 codes) and vertical directions. In addition, a temperature sensor was installed at the top of the Ophite tower, which provides temperature measurements hourly. The orientation of the building is N15\textdegree; thus, the HN3 direction (close to the north--south direction) is less exposed to the sun. A following study (Mikael et al., 2013) has given the fundamental frequencies of the building at 1.74 and 1.73~Hz for the HN2 and HN3 directions, respectively.

The natural fluctuations of the fundamental frequency for one year have been analyzed under ambient vibrations. The frequency wandering is tracked using the Random Decrement Technique (Cole, 1973), an operative and efficient method for detecting small variations with continuous recording. We observe (Fig.~\ref{fig:ophite}) that the main parameter controlling the fluctuations is the external temperature. The general observation is that frequency increases with temperature. This seems to be in contradiction with experiment performed in the laboratory (Larose et al., 2006c), where seismic velocity (or rigidity) is found to decrease with increasing temperature. Additional numerical thermo-mechanical simulations were performed, and demonstrated that the dominant effect of an external air increasing temperature is a dilation of the surface of the building (like a skin effect) that results in an increase of stress, and ultimately to an increase of rigidity (Larose and Hall, 2009).

Mikael et al. (2013) reported differences according to the building orientation. Exposure to the sun seems to have a direct effect, influencing the overall stiffness of the building by acting on the cladding or external windows. This was confirmed in refs. Clinton et al. (2006) and Herak and Herak (2010) which also reported direct correlation with rain and wind.

Although almost all previous studies showed a positive correlation between frequency and temperature, no definitive conclusions can be drawn from this analysis. As a matter of fact, the frequency versus temperature trend seems to depend on the building. The Ophite tower showed two different mechanisms, with a trend that changed between the coldest and warmest months. Additional data will be needed to draw definitive conclusions on the physical parameters explaining these trends, related to differences of the soil--structure interaction or cladding effect between each building.

\subsection{Application to a rock column}

\begin{figure}[htbp]
\centering
\includegraphics[width=0.8\textwidth]{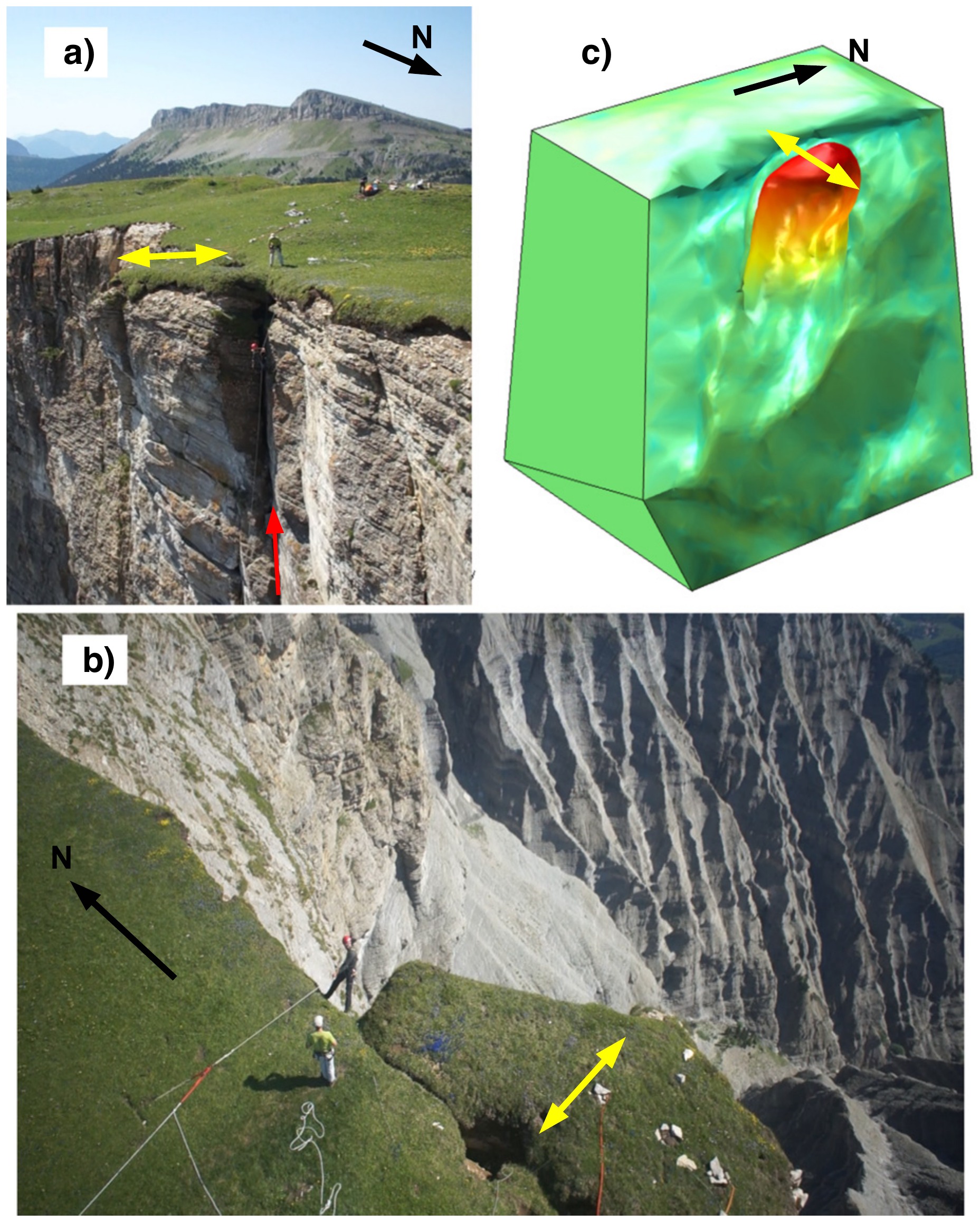}
\caption{Experimental site for monitoring the natural vibrations of a rock column, located in the south of the Vercors massif, French Alps. The black arrow indicates the North, the yellow arrow indicates the motion of the column, and the red arrow marks the rear crack. (a) and (b) pictures are taken with a drone during maintenance operations (L. Gehin/WWWprod). (c) is a numerical simulation of the column vibration amplitude in the east--west direction (yellow arrow), the topography is based on a lidar scan (details on the model are given in Bottelin et al. (2013b)). In the color scale, green corresponds to stable ground (no deformation) and red corresponds to the maximum amplitude of horizontal displacement of the natural vibration, which is of the order of 0.1~$\mu$m on the top of the column.}
\label{fig:column}
\end{figure}

\begin{figure}[htbp]
\centering
\includegraphics[width=0.6\textwidth]{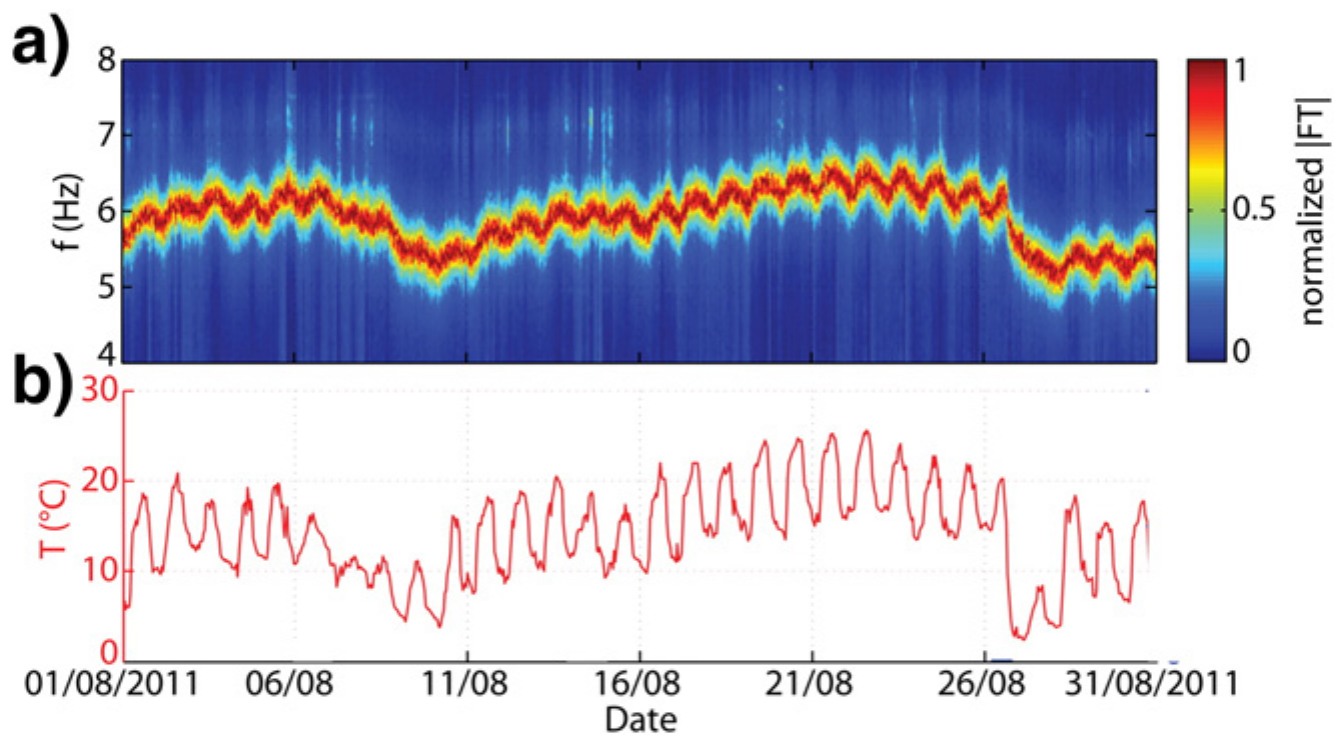}
\caption{a) Evolution of the natural resonant frequency of the rock column during summer (see Bottelin et al. (2013b) for more details on the data processing). (b) Temperature variations. Daily cycles are observable on both parameters, demonstrating the resonant frequency to depend on the external temperature.}
\label{fig:colfreq}
\end{figure}

Crystalline rocks and other rigid sediments like limestone may form high cliffs that are subject to rock falls. Erosive and complex fracturation processes can generate rock columns and flakes that progressively decouple from the mass through the breakage of rock bridges (Frayssines and Hantz, 2006). Ambient noise measurements at the top of these objects can be used to determine their natural frequencies (Bottelin et al., 2013a). To evaluate the time evolution of the natural frequencies, we deployed several three-component seismic sensors on a rock column located in the south of the French Vercors massif (see Fig.~\ref{fig:column}(a \& b)). In this figure we also display the numerical reconstruction of the deformation of the first bending mode (c). As to buildings, rock columns are subject to solar radiations and outside air temperature variations, the column was therefore also equipped with a set of thermometers. Data were continuously acquired over several seasons, which allowed us to track the frequency evolution of this first bending mode. The temporal evolution of the first natural frequency is presented in Fig.~\ref{fig:colfreq}(a), showing a perfect correlation with air temperature variations (b). Nevertheless, as to buildings, the frequency evolution does not simply relate to the evolution of the temperature in the bulk of the material. Here again, the apparent rigidity (or frequency) increases with increasing temperature. Actually, due to slow heat transfer within the column, only a thin layer undergoes daily thermal variations, which result in differential thermal-induced dilation. This dilation might have two effects: the increase of stress within the bulk of the column, and the closing of the rear fracture, both leading to an increase in the resonant frequency (Bottelin et al., 2013a, 2013b). Understanding the coupling between the natural frequency and environmental changes as temperature thus requests a precise thermo-mechanic model.

On the other hand, ambient seismic measurements may be used to study the evolution of the resonant frequency as a parameter measuring the column decoupling until its fall. Contact stiffness decreases with the breakage of rock bridges and the column natural frequencies decay with time (Lévy et al., 2010), providing a parameter characterizing the internal damage inside the rock mass. For instance, a limestone column was instrumented next to the site presented in Fig.~\ref{fig:column} with a temporary seismic array of short period seismometers from July to November 2007, two weeks prior the collapse of a 21,000~m$^3$ column (Lévy et al., 2010). The evolution of the fundamental frequency exhibits a significant drop (from 3.4~Hz to 2.6~Hz) two weeks before the fall, resulting from the breakage of rock bridges also attested by the increase in seismic activity (Lévy et al., 2010, 2011). This observation suggests that continuous monitoring of modal frequencies of rock columns might constitute a route to rockfall prediction, provided that the effects of other environmental changes, such as temperature as presented above, are properly constrained and removed.

\section{Hydrological forcing}
\label{sec:hydro}

\begin{figure}[htbp]
\centering
\includegraphics[width=0.6\textwidth]{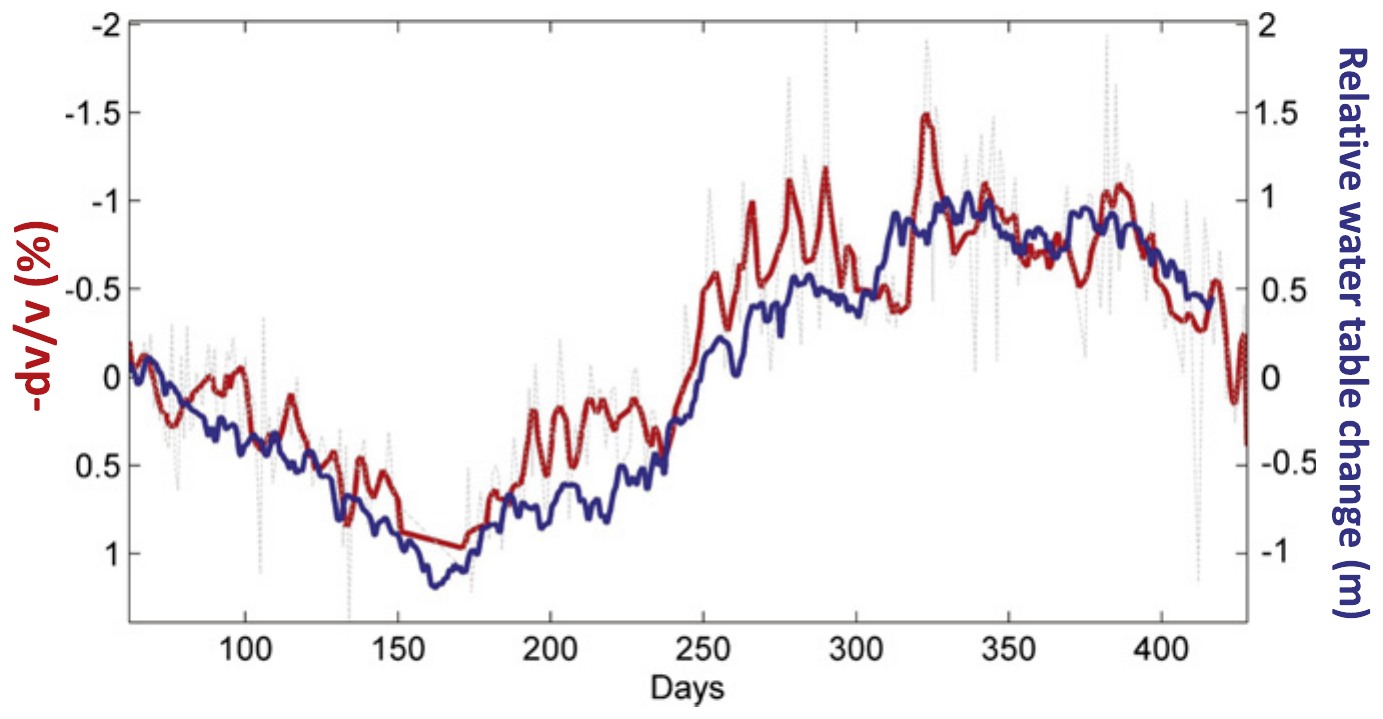}
\caption{Relative seismic velocity variations monitored at Utiku, New Zealand, in red, mostly in the 5--10~Hz range. Piezometric head levels are shown in blue. Data have been collected from late 2008 to early 2010. The relative positions of the seismic sensors were stable during the course of the experiment.}
\label{fig:utiku}
\end{figure}

The Utiku landslide (New Zealand) has been monitored over the last 40 years by the Institute of Geological and Nuclear Science (GNS Science) because its movements impair the Highway 1 integrity as well as the North Island Main Trunk railway line. Permanent GPS stations record the displacement field on a daily basis (Massey and McSaveney, 2013). They are accompanied by piezometers that are located next to the seismometers, and that record the water table level on an hourly basis. Six seismic recorders were temporarily added to the network over the September 2008 to January 2010 period, and continuously recorded the ground motion created by microtremors, trains, and over a thousand earthquakes (M 1--7.8) that occurred in the complex tectonic setting of New Zealand. During the course of the experiment, the landslide exhibited minor displacement without any periods of major acceleration. Permanent displacement during the monitoring period was slow and steady at a rate of about one meter per year. The daily cross correlation of the seismic records is an efficient tool to monitor the landslide activity. Broadband data, showing most energy in the 5--10~Hz frequency range, were correlated every day, and then the relative velocity changes were evaluated from the coda part of the correlation. Fig.~\ref{fig:utiku} (red line) presents the changes in arrival time in the coda of one pair of stations together with the water table relative charges (blue line) recorded at the lowest elevation piezometer. The similarity of both curves is striking. Other pairs of seismic stations show similar features, but with different amplitude and chronology, which seems to be in agreement with the 3D evolution of the water table within the large monitored area. The connection between water table and seismic velocity relies on the definition of the shear wave velocity: $v_s = \sqrt{\mu/\rho}$: an increase in piezometric head levels results in an increase of the density $\rho$ and in a decrease of the rigidity $\mu$, both yielding to a reduction of $v_s$. This result suggests that a 3D tracking of the underground water migration within the landslide might be possible with ambient seismic noise monitoring. Further works are currently performed to confirm this suggestion, but we can already draw the conclusion that seismic ambient noise is sensitive to hydro-meteorological conditions, and that hydrological monitoring is possible.

\section{Landslide}
\label{sec:landslide}

\begin{figure}[htbp]
\centering
\includegraphics[width=0.85\textwidth]{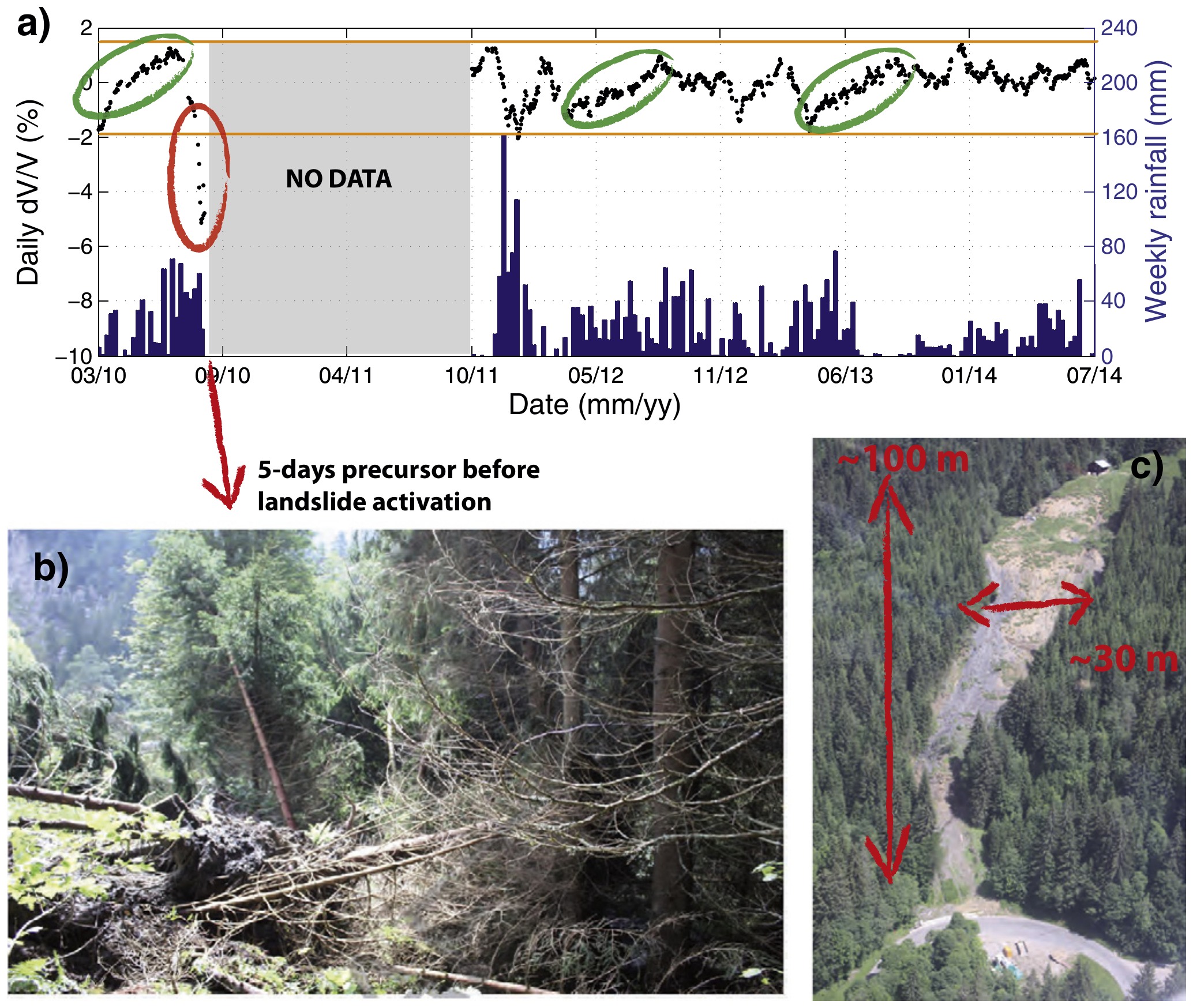}
\caption{The landslide is located in the swiss Alps, next to Les Diablerets. (a) Relative velocity changes $\dvv$ observed since January 2010, obtained from correlating continuous recordings acquired by two vertical seismic sensors across the landslide. The weekly rainfall measured by a local weather station is also presented (blue bars). For more details about the processing, the reader could refer to ref. Mainsant et al. (2012a) where the 2010 data were already presented. (b) The collapse of the landslide in August 2010 damaged the forest at the foot of the slope. (c) Geometry of the landslide before the 2010 event. Seismic sensors were placed at the end of the red horizontal arrows, about 30~m apart. The landslide is 30~m large, 300~m long and has an elevation of about 100~m.}
\label{fig:diablerets}
\end{figure}

Ambient seismic noise monitoring gives directly access to the relative velocity change within the material, which is mostly a relative velocity change for the shear wave. The source of this relative velocity change can therefore either be a change of density, or of rigidity. We thus stress that a drop in the rigidity of the material constituting a landslide could induce a drop in $\dvv$ during the slope failure. This idea was tested on a small but very active landslide located next to Les Diablerets ski resort in Switzerland. This landslide has been continuously monitored with at least two vertical seismic sensors since the beginning of 2010 (see Fig.~\ref{fig:diablerets}(c)).

Ambient seismic noise data were processed following the workflow presented in Section~\ref{sec:monitoring}, which yielded daily relative velocity changes displayed in Fig.~\ref{fig:diablerets}(a). Note that seismic data were filtered in the 8--14~Hz frequency bands, thus focusing on surface waves with depth penetration of a few meters to about 15~m. This depth window corresponds to the deforming layer constituting the landslide. In the same graph we reported weekly averaged rainfalls (blue bars). We can first point out $\dvv$ fluctuations along the years of the order of $\pm$2\% marked by two horizontal yellow lines. These fluctuations are associated to environmental changes, such as temperature, thaw/freezing cycles, snow cover and melting. In green we circle periods of slowly increasing velocities, that roughly correspond from late spring to early fall seasons, during which field observations demonstrated that the landslide material was drying (at least at the surface). Another example can be searched in late fall 2011 (right after the ``no data'' period), when strong precipitations were associated to a velocity drop (from +2\% to $-2$\% within one month). These are just illustrative examples amongst many other hydro-meteorological processes that can potentially induce $\dvv$ fluctuations. Other fluctuations along the calendar year are harder to interpret, since environmental changes sometimes result in antagonist effects on $\dvv$. Temperature, for instance, can have opposite effects: during winter season, freezing will harden the soil, when the snow cover is limited (dry snow), whereas a thicker snow layer considerably increases the soil moisture (wet snow). In that case, the effect of outside air temperature on the soil rigidity cannot be derived unless we have a precise model for heat transfer in the snow cover. Also, due to slow diffusion/migration of water in the soil, hydrological changes might have mechanical effects on the soil with various and significant delays. The general conclusion here is that, because of the structural complexity of the landslide, further studies are requested to better understand the coupling between the soil rigidity and environmental factors, which should integrate all hydro-meteorological data over a long period of time (several months at least) together with geophysical observations.

Apart from these variations along the years, there is one event that we have to focus on, where the velocity dropped far beyond the $-2$\% fluctuation limit. This drop occurred from July 15th to August 19th, 2010. In particular, the apparent velocity reduced from $-2$\% to $-7$\% from August 15th to August 19th (Mainsant et al., 2012a). This drop in rigidity resulted in a failure of the slope that developed from the 20th to the 22nd of August. A surface displacement of 20~m was observed in two days, resulting in devastating the forest at the foot of the slope (see Fig.~\ref{fig:diablerets}(b)) and the creation of a bulge threatening the road. In the present case, the relative velocity drop thus constituted a five-day precursor signal for the landslide acceleration, a mechanical behavior confirmed by laboratory experiments (Mainsant et al., 2012b). Additionally, from a spectral analysis of the velocity decrease and a standard surface wave analysis, it was possible to determine the location of the change at the base of the sliding layer: rigidity changes happened in a layer located at about 9 to 11~m depth.

During spring and summer 2010, many heavy rains were observed, associated to significant water table fluctuations. None of them but the latter triggered the slope failure, which demonstrates that hydrological data could not be used solely to predict the landslide activity.

To conclude on the previous sections, ambient seismic noise allows to monitor tiny changes in the subsurface, related for instance to thermal variations or to variations of hydrological parameters. We can track those external changes through the monitoring of the relative seismic velocity change of the waveforms, or the evolution of resonant frequencies of modal structures. In the last case corresponding to an active landslide, fluctuations of $\dvv$ due to variations of the environment seem to be restricted to $\pm$2\%, such that larger variations have to be interpreted as internal modifications, such as change in the rheology. The next section focuses on the other aspect of environmental seismology: the identification of new kinds of seismic sources hidden in the background noise, in relation to the activity of the environment.

\section{Identifying new seismic sources in the cryosphere and in the hydrosphere}
\label{sec:sources}

\subsection{Vibrations in the ice}

\begin{figure}[htbp]
\centering
\includegraphics[width=0.75\textwidth]{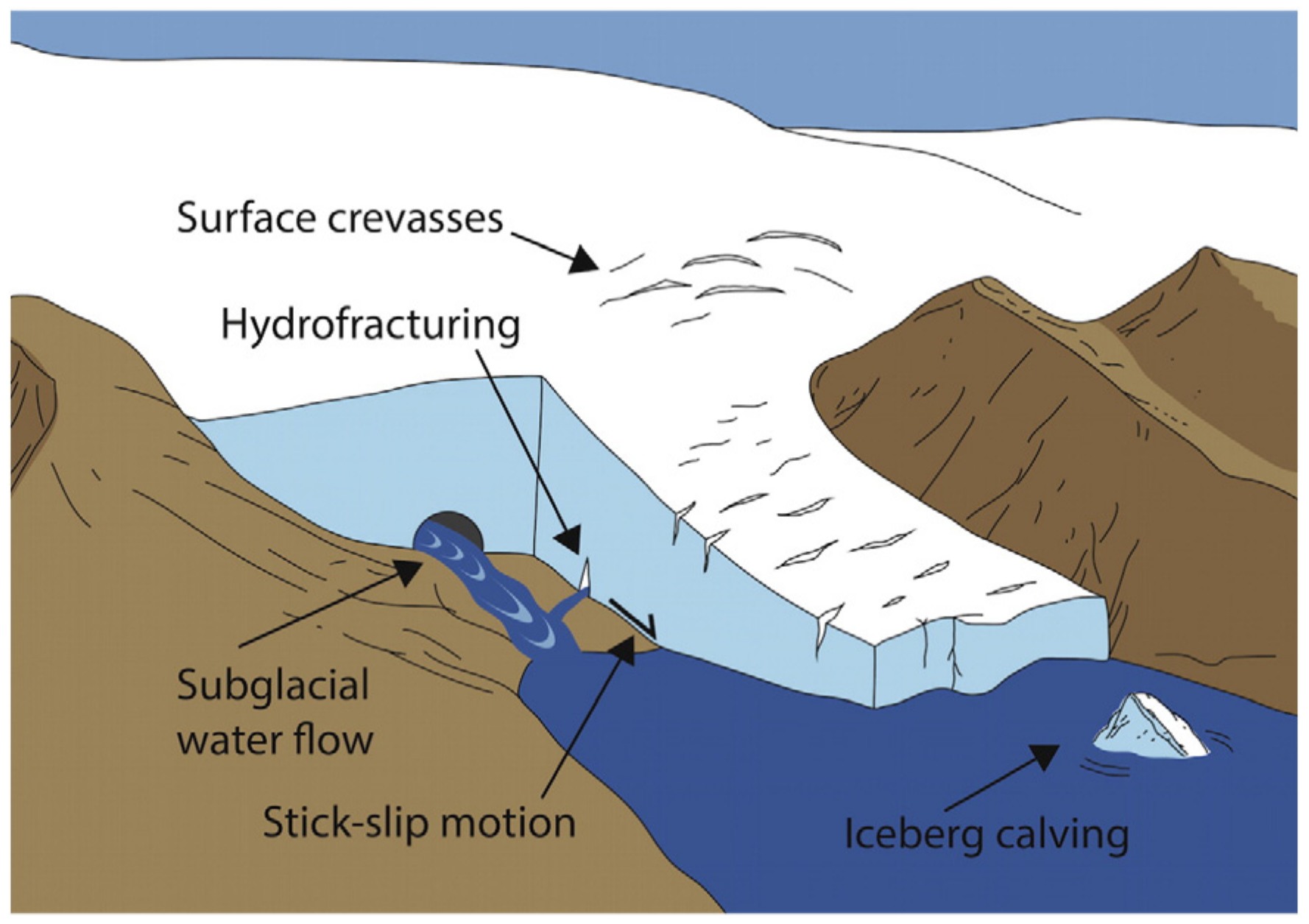}
\caption{Sketch of different sources of seismic vibrations in the ice.}
\label{fig:ice}
\end{figure}

With the advent of digital and portable instrumentation, analysis of seismic sources in glacial ice has moved into the focus of environmental seismology. Perhaps most importantly, seismic monitoring has revealed that at least part of glacier and ice stream sliding occurs as sudden stick-slip events, analogous to earthquake faulting (Allstadt and Malone, 2008; Wiens et al., 2008). Whereas seismic source studies have dominated the cryosphere component of environmental seismology, investigations of seismic wave propagation from natural seismicity are playing an increasing role, as well (Wittlinger and Farra, 2012; Harland et al., 2013). A sketch of seismic sources in the ice is proposed in Fig.~\ref{fig:ice}. Recently, this has allowed us to estimate thicknesses of floating ice, such as sea-ice (Marsan et al., 2012) and ice shelves (Zhan et al., 2014) from ambient noise.

The magnitude range of seismic sources is testimony to the wide spectrum of seismogenic processes in glaciers and ice sheets. Whereas formation and extension of surface crevasses manifests itself as events with negative magnitudes (Walter et al., 2009), iceberg detachment can produce magnitude M5 ``glacial earthquakes'', which can be detected at thousands of kilometer distances (Nettles and Ekström, 2010). Slip events beneath Antarctic ice streams are even equivalent to M7 events, however their source duration of about half an hour is responsible for relatively low seismic amplitudes (Wiens et al., 2008). Many other englacial source types of intermediate magnitudes exist, including water resonances (West et al., 2010; Röösli et al., 2014a), stick-slip tremor (MacAyeal et al., 2008; Winberry et al., 2013) and hydrofracturing (Walter et al., 2010). However, their seismic signals are often complex and embedded in strong background noise leaving many questions about source processes unanswered.

The detachment of icebergs, so-called ``calving'' events, has received particular scientific attention. Seismology offers new perspectives on remote detection and monitoring of dynamic ice discharge to the ocean. Such monitoring is desperately needed, as dynamic discharge remains the largest uncertainty in predictive ice sheet modeling (Pfeffer et al., 2008). Glacial earthquakes are typically generated when the largest icebergs detach from a grounded terminus in Greenland or Antarctica and subsequently capsize (Nettles and Ekström, 2010). Their long-period seismic surface waves can be modeled with single forces, in contrast to force couples describing tectonic earthquakes. A common explanation for the single force mechanisms are contact points with the glacier terminus or fjord bottom, across which the iceberg hinges as it capsizes (Tsai et al., 2008). Recent evidence from laboratory experiments suggests that in addition to contact forces, hydrodynamic pressure forces are also involved in glacial earthquake generation. This point has to be settled if source parameters of glacial earthquake are to be interpreted in terms of iceberg volumes (Kaluzienski et al., 2014).

Monitoring of seismic emission from basal processes is an attractive and cheap alternative to conventional glaciological techniques needed to access the glacier bed. Much consideration has been given to daily and subdaily slip events of Antarctica's Whillans Ice Stream. During these tidally modulated sliding episodes the ice stream ``leaps forward'' by tens of centimeters over a period of 10 to 30~min (Bindschadler et al., 2003). In some locations, this constitutes up to 90\% of the entire ice stream motion (Winberry et al., 2014). Besides these large scale events, ice stream beds also host smaller stick-slip events. Here, the contribution of individual events to ice stream flow is negligible (Anandakrishnan and Bentley, 1993), however, large numbers of micro faulting dislocations may combine to substantially influence basal motion. In fact, beneath the Whillans Ice Stream, slip events across a single rupture plane superimpose rapidly enough to produce gliding harmonic tremor (Winberry et al., 2013).

Accumulating evidence indicates that seismogenic stick-slip motion exists beneath glaciers and ice streams outside of Antarctica, as well (Allstadt and Malone, 2008; Röösli et al., 2014b). It is left to be shown for which types of glaciers stick-slip motion is characteristic and how small scale and large scale ruptures interact. In any case, traditional sliding theories, which describe basal motion as a combination of viscous deformation and regelation processes (Paterson and Cuffey, 1994), may have to be reconsidered if stick-slip mechanisms turn out to play a substantial and widespread role in glacier motion. On the other hand, seismogenic faulting at the base of glaciers and ice sheets offers a unique natural laboratory to study earthquake nucleation. Compared to earthquake fault zones, glacier beds can be more easily accessed. Moreover, in comparison to the Earth's crust, glacier ice is highly homogeneous and thus gives rise to cleaner seismic signals. Glacier sliding will therefore likely remain an active field of environmental seismology for years to come, whose findings may profoundly impact both ice sheet modeling and earthquake source physics.

\subsection{From oceanic to fluvial seismology}

The activity in the ocean has long been recognized to be a major source of ambient noise. In their paper in 2006, Stehly et al. (2006) mentioned the following: ``(\ldots) the mechanisms of generation of seismic noise are not the same in different period bands. At relatively short periods ($<$20~s), the two strongest peaks of the seismic noise, i.e., the primary and the secondary microseisms, are believed to be related to the interaction of the sea waves with the coast (Gutenberg, 1951). The primary microseism has periods similar to the main swell (10--20~s), while the secondary microseism that is the strongest peak in the noise spectrum originates from the nonlinear interaction between direct and reflected swell waves that results in half period (5--10~s) pressure variations (Longuet-Higgins, 1950).'' At even larger periods ($>$100~s), Rhie and Romanowicz (2004) suggested that seismic noise sources, the Earth hum, is produced by some sort of atmosphere--ocean--seafloor coupling. The mechanical coupling between the solid Earth, the oceans and the atmosphere remains subject to strong interest, and for sure observations based on the ambient seismic noise will yield to significant insights. It is nevertheless beyond the scope of the present paper to expand on these very low frequency signals. Interested readers could for example refer to Webb (1998) and Ardhuin et al. (2011) for further details.

\begin{figure}[htbp]
\centering
\includegraphics[width=0.75\textwidth]{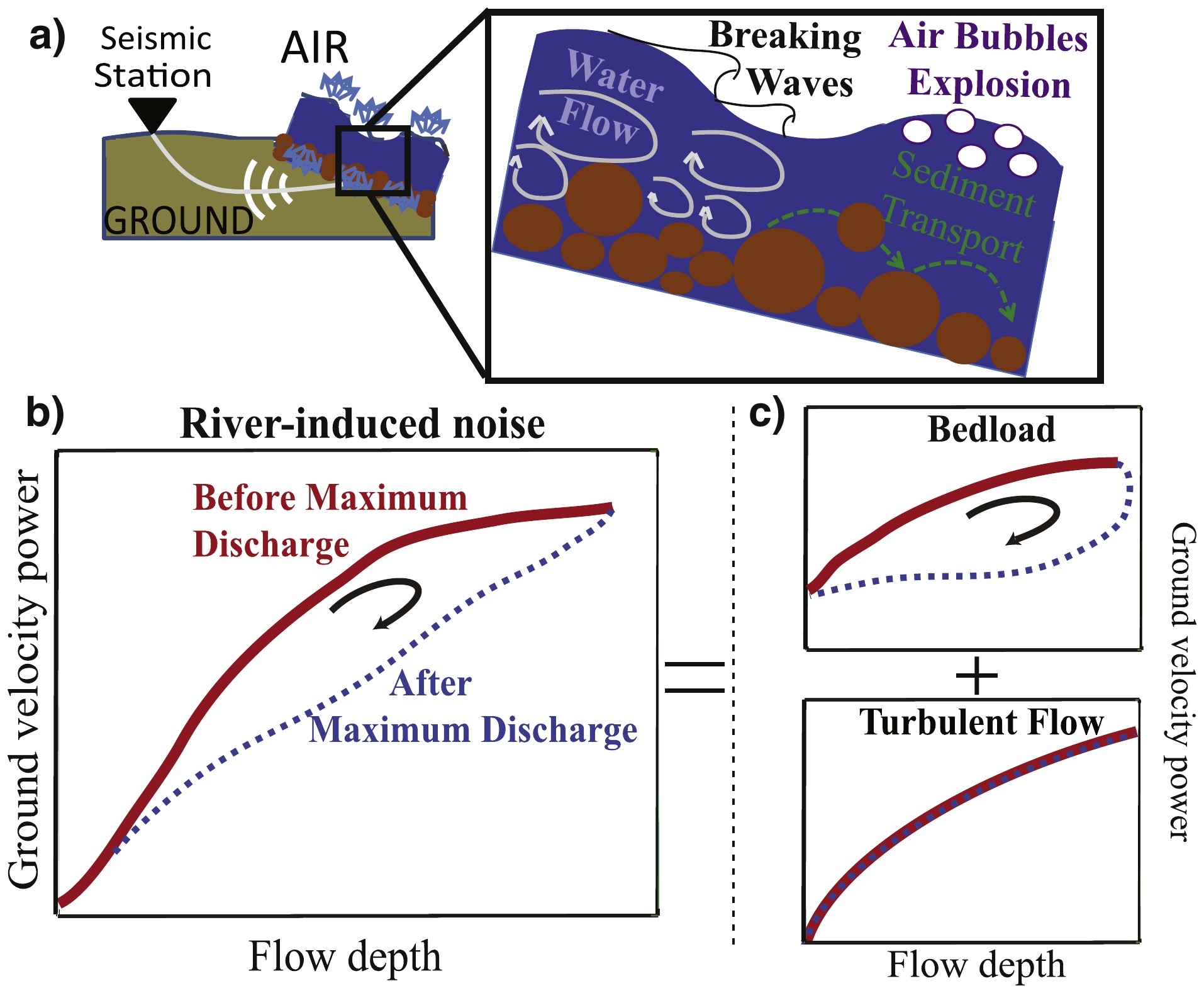}
\caption{(a) Cartoons showing (left) the deployment technique used to monitor rivers from seismic observations and (right) selected fluvial processes potentially generating ground motion. (b) Schematics of the clockwise hysteresis commonly observed in the river-induced ground power as a function of flow depth. (c) Illustration of the relative contribution of bedload-induced noise versus turbulent-flow-induced noise into the resulting hysteresis curve shown in (b). The theory proposed by Gimbert et al. (2014) allows extracting turbulent-flow-induced noise from the total river-induced noise, so that reliable estimates on the bedload flux can be obtained (Tsai et al., 2012; Gimbert et al., 2014).}
\label{fig:river}
\end{figure}

At higher frequencies that are more relevant to shallower structure studies, another kind of seismic sources should be considered, which corresponds to rivers. As anyone can easily experience, rivers generate audible noise. What about seismic noise? Depending on river flow configurations, sound and/or seismic waves may be generated by complex river processes such as the transport of sediments, the turbulent flow of water, the explosion of air bubbles, and the propagation of gravity waves or breaking waves at the river surface (see sketch in Fig.~\ref{fig:river}(a)). In recent years, numerous field and theoretical investigations have been conducted to understand which fluvial processes are mainly responsible for ground shaking, what are their specific seismic signature and how can the seismic signal be used to characterize the mechanics of the source and potentially provide new insights into the physics of fluvial processes.

By instrumenting the main stream of a small basin in northwestern Italy (the Gallina valley), Govi et al. (1993) reported a direct relationship between ground motion amplitude and water discharge. Later on, Burtin et al. (2008) provided solid evidences that a significant part of the seismic energy is attributed to bedload. By analyzing ground motion power at various places along the Trisuli River (Himalaya), they reported a seasonal variation of seismic response for a given discharge, termed hysteresis (see Fig.~\ref{fig:river}(b)). Clockwise hysteresis in ground motion power has been widely observed along various rivers and at various timescales (Hsu et al., 2011; Schmandt et al., 2013; Roth et al., 2014; Diaz et al., 2014), while counter-clockwise hysteresis is more exceptional (Diaz et al., 2014). In all cases, hysteresis in ground motion has always been attributed to hysteresis in bedload transport (see Fig.~\ref{fig:river}c), and has often been argued to result from the limited sediment supply character of the investigated rivers (Nanson, 1974; Reid et al., 1985; Whiting et al., 1999), even though other mechanisms could be at play (Roth et al., 2014).

Together with the growing interest of geomorphologists to better understand the physical processes that drive river erosion and thus landscape evolution, the appealing capability of using seismology to observe and constrain the bedload physics is undoubtedly the central motivation at the origin of the growing field of fluvial seismology. While bedload is generally the most efficient erosion mechanism (Whipple et al., 2000), bedload mechanics is poorly known. No dedicated field-measurement devices exist to monitor the key parameters used as inputs in physically-based bedload-erosion models (e.g., Sklar and Dietrich, 2004). In addition, the use of conventional techniques to estimate bedload sediment budgets (e.g., mainly sediment traps) remains very limited. These devices provide measurements with limited reliability due to 1) their invasive nature, which renders challenging their deployment during floods, and 2) the punctual measure they provide in space and time, as opposed to the intermittent character of bedload transport. Instead, the remote, non-invasive, and continuous measurements provided by seismic observations appear as a powerful tool to monitor the physics and sediment budgets associated with bedload transport. However, before useful information can be extracted from this new observational tool, appropriate theories need to be developed to isolate and quantitatively relate the contribution of the various fluvial processes into ground shaking.

\begin{figure}[htbp]
\centering
\includegraphics[width=\textwidth]{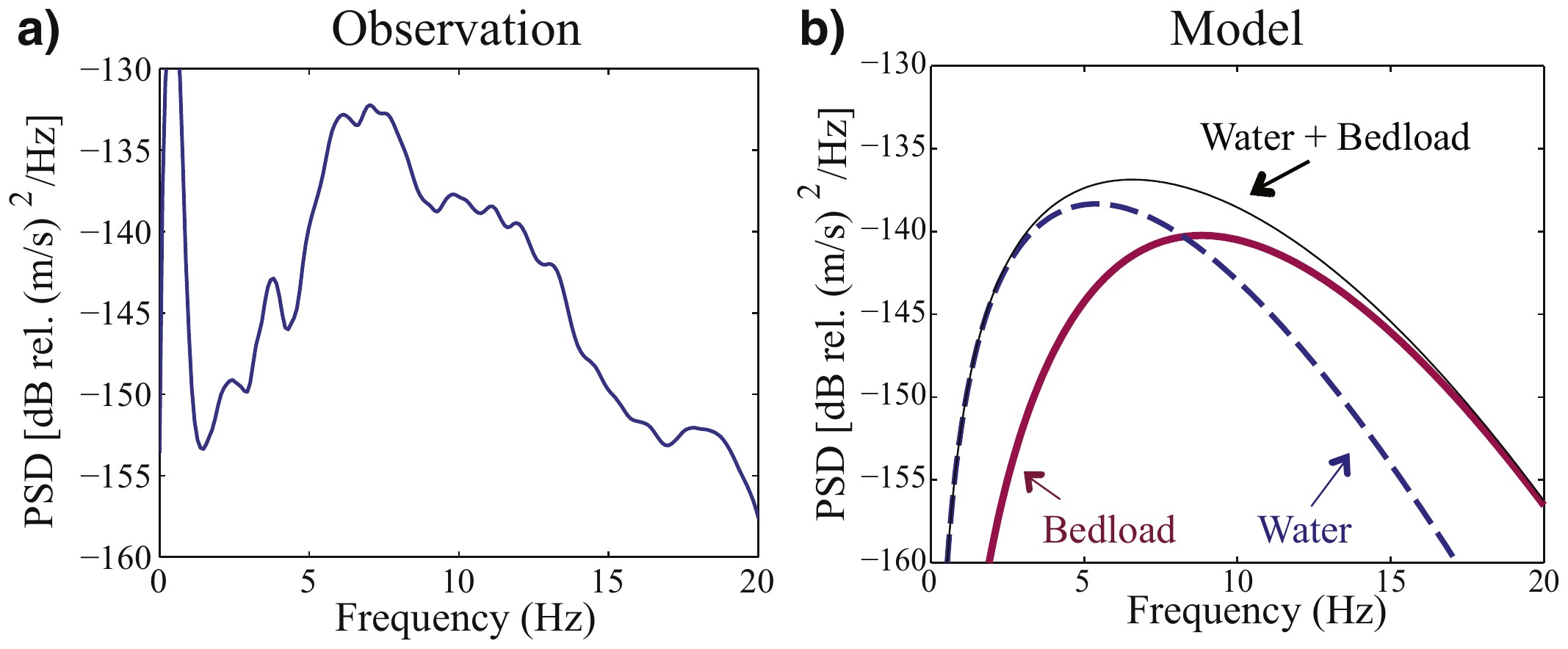}
\caption{Typical observed and modeled PSDs of ground velocity due to river-induced noise. (a) PSD observed the 15th of July 2003 at station H0460 posted about 600~m away from the Trisuli river. River flow depth was about $H$ = 3.75~m during that day (Burtin et al., 2008). (b) Model predictions for turbulent-flow-induced noise (Gimbert et al., 2014) (blue dashed line) using $H$ = 3.75~m and for bedload-induced-noise (Tsai et al., 2012) (brown continuous line) using a sediment flux $q_b$ = 0.04~m$^2$/s. The thin black line indicates the summation of both the turbulent flow and the bedload contributions. The river geometry and seismic parameters used in both the turbulent flow and the bedload model predictions are similar to Gimbert et al. (2014), and are summed up in Table~1 therein.}
\label{fig:psd}
\end{figure}

A first step towards reaching this goal has been made by Tsai et al. (2012), who proposed a theory that allows to quantitatively relate bedload to seismic ground motion. The authors used empirical formulations to describe bedload mechanics and predict seismic ground motion caused by the succession of grain impacts due to bedload transport (see Fig.~\ref{fig:psd}). They showed that the sizes of transported particles have a major control on the predicted seismic ground motion power, and that the proper knowledge of moving grain sizes allows inverting for the associated bedload flux. For an identified bedload source, this theory suggests that seismic observations can be used to quantitatively constrain the bedload physics and estimate the associated bedload flux. However, while hysteresis is an indicator of the sensitivity of seismic observations to bedload (Roth et al., 2014), other fluvial processes may potentially generate significant ground motion, and could lead to poor estimates of the bedload flux inverted from Tsai et al. (2012). In fact, observational evidences reported at the Torrent de Saint Pierre (French Alps) (Burtin et al., 2011) and at Hance Rapids of the Colorado River (United States) (Schmandt et al., 2013) suggest that water-flow can indeed significantly contribute to the river-induced seismic noise.

To allow separating the contribution of bedload and water-flow from the river-induced noise (see Fig.~\ref{fig:river}(c)), Gimbert et al. (2014) recently proposed a mechanical framework that describes seismic-noise generation by turbulent flow in rivers. The authors formulated how turbulent-flow-velocity fluctuations operating within the roughness layer of rivers generate pressure fluctuations on river-bed grains, and thus seismic waves. Their model is able to predict the raw amplitude and specific spectral signature of water-flow-induced noise previously identified at Hance Rapids (Schmandt et al., 2013), and their theory explains certain features that had previously been reported in observations, such as the lower frequency range of water-flow-induced noise as compared to bedload (see Fig.~\ref{fig:psd}), which is mainly inherited from the specific spectral scaling of flow turbulence (Kolmogorov, 1941). Gimbert and co-authors also showed that bedload and turbulent flow can independently be monitored at a given site by deploying seismic stations at various distances from the river. This suggests that reliable estimates on bedload fluxes may be obtained from seismic stations deployed relatively close to rivers, while the bed shear stress or water flow depth can be inverted from the seismic signal using seismic stations deployed farther.

These recent discoveries in the field of fluvial seismology provide solid evidences that seismic observations allow quantifying the physics involved in fluvial dynamics and landscape evolution. The remarkable capability of seismic observations to provide unique insights on the physics of the fluvial processes that operate at the grain scale together with the promising ability to estimate bedload sediment budgets from continuous measurements acquired at various temporal scales (from the storm up to the multi-seasonal scale) is very promising for future investigations in the novel, multi-disciplinary and very exciting field of fluvial seismology.

\section{Conclusion}

In this article, we reviewed different aspects of a rapidly developing field of seismology, which we can call environmental seismology. We presented a series of experiments that took benefit of ambient seismic vibrations to monitor the changes in the mechanical properties of the subsurface, or to identify new sources of seismic signals that occur outside the solid Earth.

The evolution of the mechanical properties is related either to the resonant frequencies of modal structure, or to the relative seismic velocity changes, and are attributed to external environmental changes such as temperature, soil moisture, and water table. Isolating and understanding the effect of the environment on the seismic parameters, such as rigidity, density, or damping also allows us to identify other origins of changes that can be related to internal processes. In this paper, for instance, we reported on a landslide that experienced a drop in rigidity five days prior to its failure, a rigidity modification that could not be solely attributed to environmental changes, and which constitutes a candidate for landslide failure precursor. Other recent studies have addressed the monitoring of geological structures that are deeper than those investigated in the present paper. These latter studies include monitoring and locating changes associated to an earthquake (Brenguier et al., 2008a; Obermann et al., 2014) or changes associated to forthcoming/developing volcanic eruptions (Brenguier et al., 2008b; Obermann et al., 2013). In both cases, the authors report that the modification in the seismic wave field due to mechanisms at depth are perturbed (or polluted) by changes in environmental conditions, especially from seasonal hydro-meteorological changes. In those studies, analyzing precisely the evolution of $\dvv$ in time and space allowed us to separate the reversible effect of the environment from internal modifications induced by faulting or volcanic eruptions. This sheds light on the necessity to better understand the role of the environment on the mechanical properties of the subsurface to improve our understanding of natural hazards such as landslides, volcanoes or earthquakes, and potentially improving the prediction of the formers.

In this article, we also developed on new sources of seismic signals that are related to or triggered by environmental solicitations, such as wind and storm in the atmosphere or in the ocean, water flow and sediment transport in the rivers, glaciers and ice sheets deformation and fracturation. Understanding these sources, their location, occurrence, focal mechanisms, and frequency content will shed a new light on the complex phenomena taking place at the surface of the Earth. In particular, we believe that seismic activity emitted by glaciers and ice sheets will help better understand their deformation and evolution, and also that noise emitted by rivers will help better understand and quantify river erosion. Those concerns are far beyond the scope of standard seismology.

These observations demonstrate that seismic waves, especially the ambient background noise, can be used to monitor environmental processes and their mechanical coupling with the solid Earth. In particular, this offers a new route for studying Earth surface processes, such as those developing in the hydrosphere, the cryosphere or even the atmosphere, together with better quantifying internal changes related to telluric phenomena.

\section*{Acknowledgments}

The authors acknowledge fundings from the VOR program (Université J. Fourier), the ERC Adv. Grant Whisper 227507, and the Labex OSUG@2020 (ANR10 LABX56). We acknowledge help from Peter Moore for drawing Fig.~\ref{fig:ice}.

\section*{References}
\begingroup
\small
\setstretch{1.1}
\setlength{\parindent}{-1.5em}
\setlength{\leftskip}{1.5em}
\setlength{\parskip}{0.3em}
\sloppy

Allstadt, K., Malone, S.D., 2008. Swarms of repeating stick-slip icequakes triggered by snow loading at Mount Rainier volcano. J. Geophys. Res. Earth Surf. 119 (5), 1180--1203.

Anandakrishnan, S., Bentley, C., 1993. Micro-earthquakes beneath ice streams B and C, West Antarctica: observations and implications. J. Glaciol. 31 (133), 455--462.

Ardhuin, F., Stutzmann, E., Schimmel, M., Mangeney, A., 2011. Ocean wave sources of seismic noise. J. Geophys. Res. 116, C09004.

Astiz, L., Earle, P., Shearer, P., 1996. Global stacking of broadband seismograms. Seismol. Res. Lett. 67, 8--18.

Bensen, G., Ritzwoller, M., Barmin, M., Levshin, A., Lin, F., Moschetti, M., Shapiro, N., Yang, Y., 2007. Processing seismic ambient noise data to obtain reliable broad-band surface wave dispersion measurements. Geophys. J. Int. 169, 1239--1260.

Bindschadler, R.A., King, M.A., Alley, R.B., Anandakrishnan, S., Padman, L., 2003. Tidally controlled stick-slip discharge of a West Antarctic ice. Science 301 (5636), 1087--1089.

Bottelin, P., Jongmans, D., Baillet, L., Lebourg, T., Hantz, D., Lévy, C., Le Roux, O., Cadet, H., Lorier, L., Rouiller, J., Turpin, J., Darras, L., 2013a. Spectral analysis of prone-to-fall rock compartments using ambient vibrations. J. Environ. Eng. Geophys. 18, 205--217.

Bottelin, P., Lévy, C., Baillet, L., Jongmans, D., Gueguen, P., 2013b. Modal and thermal analysis of Les Arches unstable rock column (Vercors massif, French Alps). Geophys. J. Int. 194, 849--858.

Brenguier, F., Campillo, M., Hadziioannou, C., Shapiro, N.M., Nadeau, R.M., Larose, E., 2008a. Postseismic relaxation along the San Andreas Fault at Parkfield from continuous seismological observations. Science 321 (5895), 1478--1481. \url{https://doi.org/10.1126/science.1160943}.

Brenguier, F., Shapiro, N.M., Campillo, M., Ferrazzini, V., Duputel, Z., Coutant, O., Nercessian, A., 2008b. Toward forecasting volcanic eruptions using seismic noise. Nat. Geosci. 1, 126--130. \url{https://doi.org/10.1038/ngeo104}.

Burtin, A., Bollinger, L., Vergne, J., Cattin, R., Nabelek, J.L., 2008. Spectral analysis of seismic noise induced by rivers: a new tool to monitor spatiotemporal changes in stream hydrodynamics. J. Geophys. Res. 113 (B5), B05301.

Burtin, A., Vergne, L.B.J., Cattin, R., Nabelek, J., 2011. Towards the hydrologic and bed load monitoring from high-frequency seismic noise in a braided river: the `torrent de St Pierre', French Alps. J. Hydrol. 408 (1--2), 43--53.

Campillo, M., Paul, A., 2003. Long range correlations in the diffuse seismic coda. Science 299, 547--549.

Campillo, M., Roux, P., Shapiro, N.M., 2011. Using seismic noise to image and to monitor the solid earth. In: Gupta, H.K.E. (Ed.), Encyclopedia of Solid Earth Geophysics. Springer, pp. 1230--1235.

Clinton, J., Bradford, S., Heaton, T., Favela, J., 2006. The observed wander of the natural frequencies in a structure. Bull. Seismol. Soc. Am. 96, 237--257.

Cole, H., 1973. On-line failure detection and damping measurement of aerospace structures by random decrement signatures. Technical Report, CR-2205. NASA.

Curtis, A., Gerstoft, P., Sato, H., Snieder, R., Wapenaar, K., 2006. Seismic interferometry---turning noise into signal. Lead. Edge 25, 1082--1092.

Derode, A., Larose, E., Tanter, M., de Rosny, J., Tourin, A., Campillo, M., Fink, M., 2003. Recovering the Green's function from field--field correlations in an open scattering medium. J. Acoust. Soc. Am. 113, 2973--2976.

Diaz, J., Ruiz, M., Crescentini, L., Amoruso, A., Gallart, J., 2014. Seismic monitoring of an alpine mountain river. J. Geophys. Res. 119 (2014JB010955).

Doebling, S., Farrar, C., Prime, M., Shevitz, D., 1996. Damage identification and health monitoring of structural and mechanical systems from changes in their vibration characteristics: a literature review. Technical Report, LA-13070-MS. Los Alamos National Laboratory.

Duvall, T.L., Jefferies, S.M., Harvey, J.W., Pomerantz, M.A., 1993. Time--distance helioseismology. Nature 362, 430--432.

Farrar, C., Worden, K., 2007. An introduction to structural health monitoring. Phil. Trans. R. Soc. London 365, 303--315.

Frayssines, M., Hantz, D., 2006. Failure mechanisms and triggering factors in calcareous cliffs of the subalpine ranges (French Alps). Eng. Geol. 86, 256--270.

Gimbert, F., Tsai, V.C., Lamb, M.P., 2014. A physical model for seismic noise generation by turbulent flow in rivers. J. Geophys. Res. Earth Surf. 119, 2209--2238. \url{https://doi.org/10.1002/2014JF003201}.

Govi, M., Maraga, F., Moia, F., 1993. Seismic detectors for continuous bed load monitoring in a gravel stream. Hydrol. Sci. J. 38, 123--132.

Gutenberg, B., 1951. Observation and theory of microseisms. In: Malone, T.F. (Ed.), Compendium of Meteorology. Am. Meteorol. Soc., Providence, R. I., pp. 1303--1311.

Hadziioannou, C., Larose, E., Coutant, O., Roux, P., Campillo, M., 2009. Stability of monitoring weak changes in multiply scattering media with ambient noise correlation: laboratory experiments. J. Acoust. Soc. Am. 125, 3688--3695.

Harland, S., Kendall, J.-M., Stuart, G., Lloyd, G., Baird, A., Smith, A., Pritchard, H., Brisbourne, A., 2013. Deformation in Rutford ice stream, West Antarctica: measuring shear-wave anisotropy from icequakes. Ann. Glaciol. 54 (64), 105--114.

Herak, M., Herak, D., 2010. Continuous monitoring of dynamic parameters of the DGFSM building (Zagreb, Croatia). Bull. Earthq. Eng. 8, 657--669.

Hsu, L., Finnegan, N.J., Brodsky, E.E., 2011. A seismic signature of river bedload transport during storm events. Geophys. Res. Lett. 38 (13), L13407. \url{https://doi.org/10.1029/2011GL047759}.

Humar, J.L., 1990. Dynamics of Structures. Prentice-Hall, Englewood Cliffs, NJ.

Kaluzienski, L., Burton, J., Cathles, M., 2014. Iceberg capsize hydrodynamics and the source of glacial earthquakes. Bull. Am. Phys. Soc.

Kolmogorov, A., 1941. The local structure of turbulence in incompressible viscous fluid for very large Reynolds. Proc. USSR Ac. Sci. 30, pp. 299--303.

Larose, E., Hall, S., 2009. Monitoring stress related velocity variation in concrete with a $2 \times 10^{-5}$ relative resolution using diffuse ultrasound. J. Acoust. Soc. Am. 125, 1853--1856.

Larose, E., Derode, A., Campillo, M., Fink, M., 2004. Imaging from one-bit correlations of wideband diffuse wave fields. J. Appl. Phys. 95, 8393--8399.

Larose, E., Derode, A., Clorennec, D., Margerin, L., Campillo, M., 2005. Passive retrieval of Rayleigh waves in disordered elastic media. Phys. Rev. E 72, 046607.

Larose, E., Montaldo, G., Derode, A., Campillo, M., 2006a. Passive imaging of localized reflectors and interfaces in open media. Appl. Phys. Lett. 88, 104103.

Larose, E., Margerin, L., Derode, A., van Tiggelen, B., Campillo, M., Shapiro, N., Paul, A., Stehly, L., Tanter, M., 2006b. Correlation of random wavefields: an interdisciplinary review. Geophysics 71 (4, Suppl. S), SI11--SI21. \url{https://doi.org/10.1190/1.2213356}.

Larose, E., De Rosny, J., Margerin, L., Anache, D., Gouedard, P., Campillo, M., Van Tiggelen, B., 2006c. Observation of multiple scattering of kHz vibrations in a concrete structure and application to weak changes monitoring. Phys. Rev. E 73, 16609.

Lévy, C., Baillet, L., Jongmans, D., Mourot, P., Hantz, D., 2010. Dynamic response of the Chamousset rock column (Western Alps, France). J. Geophys. Res. 115, F04043.

Lévy, C., Jongmans, D., Baillet, L., 2011. Analysis of seismic signals recorded on a prone-to-fall rock column (Vercors massif, French Alps). Geophys. J. Int. 186, 296--310.

Lobkis, O.I., Weaver, R.L., 2003. Coda-wave interferometry in finite solids: recovery of P-to-S conversion rates in an elastodynamic billiard. Phys. Rev. Lett. 90, 254302.

Longuet-Higgins, M., 1950. A theory of the origin of microseisms. Philos. Trans. R. Soc. Lond. 243, 137--171.

MacAyeal, D., Okal, E., Aster, R., Bassis, J., 2008. Seismic and hydroacoustic tremor generated by colliding icebergs. J. Geophys. Res. Earth Surf. (F3). \url{https://doi.org/10.1029/2008JF001005}.

Mainsant, G., Larose, E., Broennimann, C., Michoud, C., Jongmans, D., 2012a. Ambient seismic noise monitoring of a clay landslide: toward failure prediction. JGR 39, L19301.

Mainsant, G., Jongmans, D., Chambon, G., Larose, E., Baillet, L., 2012b. Shear-wave velocity as an indicator for rheological changes in clay materials: lessons from laboratory experiments. J. Geophys. Res. 39, L19301.

Marsan, D., Weiss, J., Larose, E., Métaxian, J.-P., 2012. Sea-ice thickness measurement based on the dispersion of ice swell. J. Acoust. Soc. Am. 131, 80.

Massey, C.I., McSaveney, D.N.P.M.J., 2013. Patterns of movement in reactivated landslides. Eng. Geol. 159, 1--19.

Michel, C., Gueguen, P., 2010. Time--frequency analysis of small frequency variations in civil engineering structures under weak and strong motions using a reassignment method. Struct. Control. Health Monit. 9, 159--171.

Mikael, A., Gueguen, P., Bard, P.-Y., Roux, P., Langlais, M., 2013. The analysis of long-term frequency and damping wandering in buildings using the random decrement technique. Bull. Seismol. Soc. Am. 103, 236--246.

Nanson, G.C., 1974. Bedload and suspended-load transport in a small, steep, mountain stream. Am. J. Sci. 274 (5), 471--486. \url{https://doi.org/10.2475/ajs.274.5.471}.

Nettles, M., Ekström, G., 2010. Glacial earthquakes in Greenland and Antarctica. Annu. Rev. Earth Planet. Sci. 38 (1), 467.

Obermann, A., Planès, T., Larose, E., Campillo, M., 2013. Imaging preeruptive and coeruptive structural and mechanical changes of a volcano with ambient seismic noise. J. Geophys. Res. 118, 1--11.

Obermann, A., Froment, B., Campillo, M., Larose, E., Planès, T., Valette, B., Chen, J.H., Liu, Q.Y., 2014. Seismic noise correlations to image structural and mechanical changes associated with the Mw 7.9 2008 Wenchuan earthquake. J. Geophys. Res. 119 (4), 3155--3168.

Paterson, W.S.B., Cuffey, K.M., 1994. The Physics of Glaciers vol. 3. Pergamon, Oxford.

Paul, A., Campillo, M., Margerin, L., Larose, E., Derode, A., 2005. Empirical synthesis of time-asymmetrical Green functions from the correlation of coda waves. J. Geophys. Res. 110, B08302.

Péquegnat, C., Guéguen, P., Hatzfeld, D., Langlais, M., 2008. The French accelerometric network (RAP) and national data centre (RAP-NDC). Seismol. Res. Lett. 79 (1), 79--89.

Pfeffer, W.T., Harper, J., O'Neel, S., 2008. Kinematic constraints on glacier contributions to 21st-century sea-level rise. Science 321 (5894), 1340--1343.

Planes, T., Larose, E., 2013. A review of ultrasonic Coda Wave Interferometry in concrete. Cem. Concr. Res. 53, 248--255.

Poupinet, G., Ellsworth, W.L., Frechet, J., 1984. Monitoring velocity variations in the crust using earthquake doublets: an application to the Calaveras fault, California. J. Geophys. Res. 89, 5719--5731.

Reid, I., Frostick, L.E., Layman, J.T., 1985. The incidence and nature of bedload transport during flood flows in coarse-grained alluvial channels. Earth Surf. Process. Landf. 10 (1), 33--44. \url{https://doi.org/10.1002/esp.3290100107}.

Rhie, J., Romanowicz, B., 2004. Excitation of earth's continuous free oscillations by atmosphere--ocean--seafloor coupling. Nature 431, 552--556.

Röösli, C., Walter, F., Kissling, E., Helmstetter, A., Lüthi, M., 2014a. Seismic signals associated with basal processes of the Greenland ice sheet. EGU General Assembly Conference Abstracts vol. 16, p. 6822.

Röösli, C., Walter, F., Husen, S., Andrews, L.C., Lüthi, M.P., Catania, G.A., Kissling, E., 2014b. Sustained seismic tremors and icequakes detected in the ablation zone of the Greenland ice sheet. J. Glaciol. 60 (221), 563--575.

Roth, D., 2014. Migration of a coarse fluvial sediment pulse detected by hysteresis in bedload generated seismic waves. Earth Planet. Sci. Lett. 404, 144--153.

Roux, P., Kuperman, W., the NPAL Group, 2004. Extracting coherent wave fronts from acoustic ambient noise in the ocean. J. Acoust. Soc. Am. 116, 1995--2003.

Sabra, K.G., Roux, P., Kuperman, W.A., 2005. Arrival-time structure of the time-averaged ambient noise cross-correlation function in an oceanic waveguide. J. Acoust. Soc. Am. 117, 164--174.

Sato, H., 2009. Green function retrieval from the CCF of coda waves in a scattering medium. Geophys. J. Int. 179, 1580--1583.

Schmandt, B., Aster, R.C., Scherler, D., Tsai, V.C., Karlstrom, K.K., 2013. Multiple fluvial processes detected by riverside seismic and infrasound monitoring of a controlled flood in the Grand Canyon. Geophys. Res. Lett. 40, 4858--4863.

Sens-Schönfelder, C., Larose, E., 2008. Temporal changes in the lunar soil from correlation of diffuse vibrations. Phys. Rev. E 78 (4, Part 2), 045601. \url{https://doi.org/10.1103/PhysRevE.78.045601}.

Sens-Schönfelder, C., Wegler, U.C., 2006. Passive image interferometry and seasonal variations of seismic velocities at Merapi volcano, Indonesia. Geophys. Res. Lett. 33, L21302. \url{https://doi.org/10.1029/2006GL027797}.

Shapiro, N.M., Campillo, M., 2004. Emergence of broadband Rayleigh waves from correlations of the ambient seismic noise. Geophys. Res. Lett. 31, L07614.

Shearer, P., 2009. Introduction to Seismology. Second Ed. Cambridge University Press.

Sklar, L., Dietrich, W., 2004. A mechanistic model for river incision into bedrock by saltating bed load. Water Resour. Res. 40, W06301.

Snieder, R., Larose, E., 2013. Extracting the earth response from noise measurements. Annu. Rev. Earth Planet. Sci. 41, 9.1--9.24.

Snieder, R., Safak, E., 2006. Extracting the building response using seismic interferometry: theory and application to the Millikan library in Pasadena, California. Bull. Seismol. Soc. Am. 96, 586--598.

Snieder, R., Grêt, A., Douma, H., Scales, J., 2002. Coda wave interferometry for estimating nonlinear behavior in seismic velocity. Science 295, 2253--2255.

Stehly, L., Shapiro, N., Campillo, M., 2006. A study of the seismic noise from its long-range correlation properties. J. Geophys. Res. 111, B10306.

Tsai, V.C., 2011. A model for seasonal changes in GPS positions and seismic wave speeds due to thermoelastic and hydrologic variations. J. Geophys. Res. 116, B04404.

Tsai, V.C., Minchew, B., Lamb, M.P., Ampuero, J.-P., 2012. A physical model for seismic noise generation from sediment transport in rivers. Geophys. Res. Lett. 39, L02404.

Tsai, V.C., Rice, J.R., Fahnestock, M., 2008. Possible mechanisms for glacial earthquakes. J. Geophys. Res. 113 (F03014). \url{https://doi.org/10.1029/2007JF000944}.

Walter, F., Clinton, J.F., Deichmann, N., Dreger, D.S., Minson, S.E., Funk, M., 2009. Moment tensor inversions of icequakes on Gornergletscher, Switzerland. Bull. Seismol. Soc. Am. 99 (2A), 852--870.

Walter, F., Dreger, D.S., Clinton, J.F., Deichmann, N., Funk, M., 2010. Evidence for near-horizontal tensile faulting at the base of Gornergletscher, a Swiss alpine glacier. Bull. Seismol. Soc. Am. 100 (2), 458--472.

Weaver, R.L., Lobkis, O.I., 2001. Ultrasonics without a source: thermal fluctuation correlations at MHz frequencies. Phys. Rev. Lett. 87, 134301.

Weaver, R.L., Hadziioannou, C., Larose, E., Campillo, M., 2011. On the precision of noise correlation interferometry. Geophys. J. Int. 185, 1384--1392.

Webb, S.C., 1998. Broadband seismology and noise under the ocean. Rev. Geophys. 36, 105--142.

West, M.E., Larsen, C.F., Truffer, M., O'Neel, S., LeBlanc, L., 2010. Glacier microseismicity. Geology 38 (4), 319--322.

Whipple, K.X., Hancock, G.S., Anderson, R.S., 2000. River incision into bedrock: mechanics and relative efficacy of plucking, abrasion, and cavitation. Geol. Soc. Am. Bull. 112 (3), 490--503.

Whiting, P.J., Stamm, J.F., Moog, D.B., Orndorff, R.L., 1999. Sediment-transporting flows in headwater streams. Geol. Soc. Am. Bull. 111 (3), 450--466.

Wiens, D.A., Anandakrishnan, S., Winberry, J.P., King, M.A., 2008. Simultaneous teleseismic and geodetic observations of the stick-slip motion of an antarctic ice stream. Nature 453 (7196), 770--774.

Winberry, J.P., Anandakrishnan, S., Wiens, D., Alley, R., 2013. Nucleation and seismic tremor associated with the glacial earthquakes of Whillans Ice Stream, Antarctica. Geophys. Res. Lett. 40 (2), 312--315.

Winberry, J.P., Anandakrishnan, S., Alley, R.B., Wiens, D.A., Pratt, M.J., 2014. Tidal pacing, skipped slips and the slowdown of Whillans Ice Stream, Antarctica. J. Glaciol. 60 (222), 795.

Wittlinger, G., Farra, V., 2012. Observation of low shear wave velocity at the base of the polar ice sheets: evidence for enhanced anisotropy. Geophys. J. Int. 190 (1), 391--405.

Zhan, Z., Tsai, V.C., Jackson, J.M., Helmberger, D., 2014. Ambient noise correlation on the Amery ice shelf, East Antarctica. Geophys. J. Int. 196, 1796--1802. \url{https://doi.org/10.1093/gji/ggt488}.

\endgroup

\end{document}